# Deep Neural Network Computes Electron Densities and Energies of a Large Set of Organic Molecules Faster than Density Functional Theory (DFT)

*Anton V. Sinitskiy, Vijay S. Pande*

*Department of Bioengineering, Stanford University, Stanford CA 94305*

Density functional theory (DFT) is one of the main methods in Quantum Chemistry that offers an attractive trade off between the cost and accuracy of quantum chemical computations. The electron density plays a key role in DFT. In this work, we explore whether machine learning – more specifically, deep neural networks (DNNs) – can be trained to predict electron densities faster than DFT. First, we choose a practically efficient combination of a DFT functional and a basis set (PBE0/pcS-3) and use it to generate a database of DFT solutions for more than 133,000 organic molecules from a previously published database QM9. Next, we train a DNN to predict electron densities and energies of such molecules. The only input to the DNN is an approximate electron density computed with a cheap quantum chemical method in a small basis set (HF/cc-VDZ). We demonstrate that the DNN successfully learns differences in the electron densities arising both from electron correlation and small basis set artifacts in the HF computations. All qualitative features in density differences, including local minima on lone pairs, local maxima on nuclei, toroidal shapes around C – H and C – C bonds, complex shapes around aromatic and cyclopropane rings and CN group, etc. are captured by the DNN. Accuracy of energy predictions by the DNN is ~ 1 kcal/mol, on par with other models reported in the literature, while those models do not predict the electron density. Computations with the DNN, including HF computations, take much less time that DFT computations (by a factor of ~20-30 for most QM9 molecules in the current version, and it is clear how it could be further improved).

**Introduction**

Density functional theory (DFT) is one of the main methods in Quantum Chemistry. DFT has been widely used by the scientific community because it offers a favorable trade off between the cost of computations and the accuracy of results.[1-5] Though DFT is typically faster than *ab initio* quantum chemical methods, such as the second-order Møller–Plesset perturbation theory (MP2) or coupled cluster method with single and double excitations (CCSD), modeling a single conformation of an organic molecule with nine heavy (non-hydrogen) atoms using a relatively accurate DFT functional and a large basis set may still take hours, limiting its applicability of large molecular systems.

The concept of one-particle electron density $\rho(\mathbf{r})$ plays a central role in DFT. In general, a quantum mechanical system can be characterized by a wavefunction $\Psi$. An electronic wavefunction of a molecule with $N$ electrons $\Psi(\mathbf{r}_1, \ldots, \mathbf{r}_N, \tau_1, \ldots, \tau_N)$ depends on $3N$ spatial coordinates $\mathbf{r}_1, \ldots, \mathbf{r}_N$ and $N$ spin coordinates $\tau_1, \ldots, \tau_N$. For molecular systems of practical interest, $N \gg 1$, making it difficult to work with wavefunctions because of their high dimensionality. On the contrary, the electron density, defined as

$$\rho(\mathbf{r}) = N \sum_{\tau_1} \ldots \sum_{\tau_N} \int d\mathbf{r}_2 \ldots \int d\mathbf{r}_N \left| \Psi(\mathbf{r}_1, \ldots, \mathbf{r}_N, \tau_1, \ldots, \tau_N) \right|^2, \quad (1)$$

is a function of only a three-dimensional vector, and hence is much easier to deal with from various viewpoints, including intuitive appreciation, graphical visualization and basis set expansions. This function shows how electrons are delocalized in a molecule, namely, where the probability to find an electron per unit volume is higher, and where it is lower. According to the Hohenberg–Kohn theorems,[1] the ground state energy $E$ of a molecular system can be found by a minimization of a density functional $E[n(\mathbf{r})]$; the density $n$ at which the minimum is achieved yields the ground state electron density:

$$E = \min_{n(\mathbf{r})} E[n(\mathbf{r})], \quad \rho(\mathbf{r}) = \arg\min_{n(\mathbf{r})} E[n(\mathbf{r})]. \quad (2)$$

An exact expression for the density functional $E[n(\mathbf{r})]$ is not known. Numerous approximate functional have been proposed, leading to multiple versions of DFT with various accuracy level and computational cost.

Despite the importance of electron density to DFT, very few works have directly measured the actual accuracy of DFT in predicting electron densities of molecules.[6] One of the reasons is that 'exact' electron densities – that is, electron densities guaranteed to be much more accurate than those predicted from DFT, obtained either from high-level quantum chemical computations with large basis sets, or from experimental measurements, and to be used for benchmarking as reference electron densities – are much more difficult to get than 'exact' energies. However, such benchmark studies, even less representative than benchmark studies for energies, could be important for understanding the benefits and drawback of DFT, relative comparisons of various DFT functionals, as well as comparisons of DFT to other quantum chemical methods.

Recently, machine learning (ML) – and, more specifically, deep learning, that is ML with deep neural networks (DNNs) – have achieved impressive results in dealing with large sets of data, including three-dimensional shapes and functions.[7-10] ML is also starting to be actively and fruitfully applied to various problems in Quantum Chemistry (for reviews, see Refs. 11-15). In particular, DNNs and other ML models were trained to predict energies of molecules (primarily organic) as functions of atomic coordinates with thermochemical accuracy and at much smaller computational cost than quantum chemical methods (ANI-1* potentials and many others).[16-42] However, electron densities have been out of focus of ML in this field so far.

In this work, we aim to find out whether deep learning can predict electron densities of molecules, and if so, how its performance (in terms of accuracy and speed of computations) relates to that of DFT. To the best of our knowledge, ML have been used so far to predict electron densities only in specific molecular systems, such as sulfur-crosslinked carbon nanotubes,[21] Ni/Al alloys,[22] dihydrogen, water, benzene, ethane, and malonaldehyde.[16] The question of feasibility of electron density learning in wide classes of chemical compounds remains open.

Out of several quantum chemical databases of molecules, we chose to work with QM9.[43] This is a database of 133 885 organic molecules made up of C, H, N, O and F elements. All molecules in QM9 have up to nine heavy (non-hydrogen) atoms, zero electric charge and spin. The QM9 database includes the geometries of all molecules optimized with B3LYP DFT functional and 6-31G(2df,p) basis set. Unlike other existing quantum chemical databases,[18,44-46] QM9 is at the same time easy to work with (all data files can be easily downloaded and processed locally), contains more molecules than some other databases, but still a manageable number of molecules to run quantum chemical computations for all of them with our resources, includes F (not only C, H, N, O), and focuses on conformations close to equilibrium (we chose to investigate how to do ML of electron densities in molecules near equilibrium before proceeding to non-equilibrium geometries). In this work, we report the results of ML of electron densities in the molecules from the QM9 dataset. We also make the fchk output files of our quantum chemical computations for the QM9 dataset available to all researchers by publishing it in a public repository, hoping that these data on the wavefunctions and electron densities of more than 133 thousand molecules will foster further progress in the field of ML in Quantum Chemistry.

**Results**

DFT functional PBE0 with basis set pcS-3 is practically efficient for generating a large dataset of electron densities of organic molecules

We started by choosing a DFT functional and a basis set to be used for making a database of electron densities for ML. This choice is based on two considerations: First, the chosen DFT functional with the chosen basis set should provide relatively high accuracy in the computed electron densities, in comparison to other DFT functionals and basis sets. Second, computations should be fast enough to treat ~133K molecules in a reasonable amount of time (up to several months), given the computational resources we had.

As a part of addressing the first question, we computed electron densities of the first six molecules from the QM9 database ($CH_4$, $NH_3$, $H_2O$, $C_2H_2$, HCN, $CH_2O$) with a high level *ab initio* quantum chemical method (CCSD, without a frozen core approximation) often used as a source of 'exact' wavefunctions, in a large basis set (cc-pCV5Z). Rapid increase of the cost of computations with the molecule size prevents us from carrying out such computations for much more than a few first entries in the database. For the same six molecules, we also computed electron densities using various combinations of DFT functionals and basis sets. In total, 2643 combinations were screened (excluding combinations for which computations did not converge for more than two molecules). For each DFT functional, basis set and molecule, an L1 measure of the accuracy of the predicted electron density $\rho(\mathbf{r})$ was calculated:

$$L1 = \int dV \left| \rho(\mathbf{r}) - \rho_{ref}(\mathbf{r}) \right|, \qquad (3)$$

where $\rho(\mathbf{r})$ is the evaluated electron density, $\rho_{ref}(\mathbf{r})$ is the reference ('exact', CCSD/cc-pCV5Z) electron density, and $V$ is for Vendetta (for details, see Methods). The combinations of DFT functionals and basis sets were sorted by the average value of L1 over six molecules. The top of the list is given in Table 1. From the analysis of the list we conclude that:

- combinations of DFT functionals and basis sets with low values of L1 for some of these six molecules tend to have low values of L1 for the other molecules, and vice versa;
- the top of the list is dominated by two families of functionals (ωB97X-D[47] and related,[48] and PBE0[49,50] and related), and one family of basis sets (pcS sets[51]);
- pcS-3-level basis sets yield values of L1 as low as V5Z-level basis sets do; neither a shift to pcS-4-level basis sets, nor augmentation of pcS-3-level basis sets significantly improve L1.

These results on L1 measures of the accuracy of electron densities cannot be considered as a benchmark study because of a small number of included molecules, which is a consequence of the use of a high-level *ab initio* method and a large basis set for computing the reference electron densities. However, our major conclusions from this analysis agree with some other results and theoretical considerations in the literature. In particular, a recent study compared the performance of several DFT functionals in predicting electron densities of 30 organic molecules with two to ten heavy atoms (only aug-cc-pVTZ basis set was used). The best method was found to be TPSSh, with PBE0 being only slightly worse.[52] Another study compared the performance of numerous DFT functionals over a set of 14 atoms and monoatomic ions, with a different measure of the accuracy in predicting electron densities (based on radial distribution functions). PBE0 was found to be one of the most accurate DFT methods.[6] As for the pcS family basis sets, they are rarely included into comparisons, but when they are, usually they perform well.[53-55] Finally, there are some theoretical considerations supporting that ωB97X, PBE0 and related functionals and pcS basis sets may be good candidates for generating datasets of electron densities. PBE0 functional is based to a significant degree on exact theoretical results on the energy functional, rather than fitting to experimental data.[6,49,50] ωB97X and related functionals include a relatively small number of fitted parameters.[5] pcS basis sets, unlike most other basis set families, were designed to fit experimentally measured nuclear magnetic shielding constants, not energies,[51] which may be the reason for their good performance in approximating electron densities.

To compare different DFT methods from the viewpoint of computational cost, we took 176 combinations of DFT functionals and basis sets with the lowest average values of L1, and used them to compute electron densities and energies of QM9 entries 8 000, 16 000 and 32 000. It is often stated that one needs tens of thousands of datapoints for successful ML, which guided our choice of these QM9 entries. We computed an aggregated indicator showing how much wallclock time is required to run DFT computations for a molecule of this size on a single GPU (further called 'effective time'; for exact definition and computation details, see Methods). The plot for the computational cost (estimated by effective time) vs. accuracy (estimated by average L1 over the first six QM9 molecules) shows that some combinations of a DFT functional and a basis set have a favorable tradeoff between the cost and accuracy (Fig. 1). PBE0/pcS-3, which is #4 on the list in Table 1, looks particularly attractive. Other good choices might be PBE1hPBE/pcS-3, PBE0/Def2-QZVP, and possibly ωB97X-D/cc-pV5Z (though the last combination is more expensive). As for the first three combinations on the list in Table 1, computations for ωB97X-D3[48] with pcS-3-level basis sets turned out to be prohibitively expensive, while OHSE2PBE (also known as HSE03) is less used in the literature than PBE0, and its gain in accuracy over PBE0/pcS-3 is marginal.

**Table 1**. L1 measures of the difference between 3D electron densities predicted by various DFT methods with various basis sets, and 'exact' (CCSD/cc-pCV5Z) electron densities [eq. (3)]. Data given for the first six molecules in the QM9 database. Top 20 combinations of functionals and basis sets, in the order of increasing average value of L1, are shown. Also, some other functionals are given for comparison, each with the basis set providing the highest average L1 value for this functional. Double horizontal lines show that some rows were omitted. DFT functionals are named by the corresponding keywords from Gaussian or Q-Chem, except for PBE0 (Gaussian keyword 'PBE1PBE').

| DFT functional | basis set | 1 | 2 | 3 | 4 | 5 | 6 | average |
|---|---|---|---|---|---|---|---|---|
| ωB97X-D3 | pcS-3 | 0.045 | 0.041 | 0.039 | 0.043 | 0.035 | 0.058 | 0.043 |
| ωB97X-D3 | aug-pcS-3 | 0.045 | 0.042 | 0.041 | 0.043 | 0.036 | 0.059 | 0.044 |
| OHSE2PBE | pcS-3 | 0.052 | 0.042 | 0.040 | 0.041 | 0.034 | 0.071 | 0.047 |
| PBE0 | pcS-3 | 0.058 | 0.043 | 0.039 | 0.039 | 0.034 | 0.071 | 0.047 |
| OHSE2PBE | aug-pcS-3 | 0.052 | 0.043 | 0.041 | 0.041 | 0.034 | 0.071 | 0.047 |
| ωB97X-D3 | cc-pV5Z | 0.049 | 0.040 | 0.035 | 0.053 | 0.044 | 0.064 | 0.048 |
| PBE0 | aug-pcS-3 | 0.058 | 0.044 | 0.040 | 0.040 | 0.034 | 0.071 | 0.048 |
| ωB97X-D | pcS-3 | 0.048 | 0.047 | 0.042 | 0.049 | 0.039 | 0.067 | 0.049 |
| PBEh1PBE | pcS-3 | 0.057 | 0.043 | 0.041 | 0.043 | 0.036 | 0.072 | 0.049 |
| OHSE1PBE | pcS-3 | 0.057 | 0.043 | 0.041 | 0.044 | 0.036 | 0.073 | 0.049 |
| HSEH1PBE | pcS-3 | 0.057 | 0.043 | 0.041 | 0.044 | 0.036 | 0.073 | 0.049 |
| PBEh1PBE | aug-pcS-3 | 0.057 | 0.044 | 0.042 | 0.043 | 0.036 | 0.073 | 0.049 |
| OHSE1PBE | aug-pcS-3 | 0.057 | 0.045 | 0.042 | 0.044 | 0.036 | 0.073 | 0.050 |
| HSEH1PBE | aug-pcS-3 | 0.057 | 0.045 | 0.042 | 0.044 | 0.036 | 0.073 | 0.050 |
| ωB97X-D | aug-pcS-3 | 0.049 | 0.048 | 0.044 | 0.049 | 0.040 | 0.068 | 0.050 |
| revPBE0 | pcS-3 | 0.050 | 0.050 | 0.045 | 0.041 | 0.041 | 0.072 | 0.050 |
| ωB97X-D3 | pcS-4 | 0.048 | 0.046 | 0.045 | 0.051 | 0.044 | 0.065 | 0.050 |
| ωB97X-D3 | aug-pcS-4 | 0.049 | 0.046 | 0.045 | 0.051 | 0.045 | 0.065 | 0.050 |
| revPBE0 | aug-pcS-3 | 0.051 | 0.051 | 0.046 | 0.042 | 0.041 | 0.073 | 0.051 |
| ωB97X-D3 | pc-4 | 0.049 | 0.046 | 0.046 | 0.053 | 0.046 | 0.066 | 0.051 |
| ωB97X-V | pcS-3 | 0.048 | 0.045 | 0.047 | 0.058 | 0.052 | 0.062 | 0.052 |
| APFD | cc-pV5Z | 0.058 | 0.040 | 0.037 | 0.053 | 0.046 | 0.081 | 0.053 |
| B3PW91 | pcS-3 | 0.055 | 0.050 | 0.049 | 0.052 | 0.044 | 0.085 | 0.056 |
| ωB97X | pcS-3 | 0.053 | 0.049 | 0.049 | 0.063 | 0.055 | 0.071 | 0.057 |
| mPW3PBE | cc-pV5Z | 0.064 | 0.047 | 0.044 | 0.066 | 0.057 | 0.093 | 0.062 |
| TPSSh | pcS-3 | 0.063 | 0.066 | 0.064 | 0.051 | 0.051 | 0.086 | 0.064 |
| M062X | Apr-cc-pVQZ | 0.082 | 0.062 | 0.047 | 0.079 | 0.072 | 0.092 | 0.072 |
| B3LYP | pcS-3 | 0.093 | 0.093 | 0.087 | 0.117 | 0.102 | 0.127 | 0.103 |

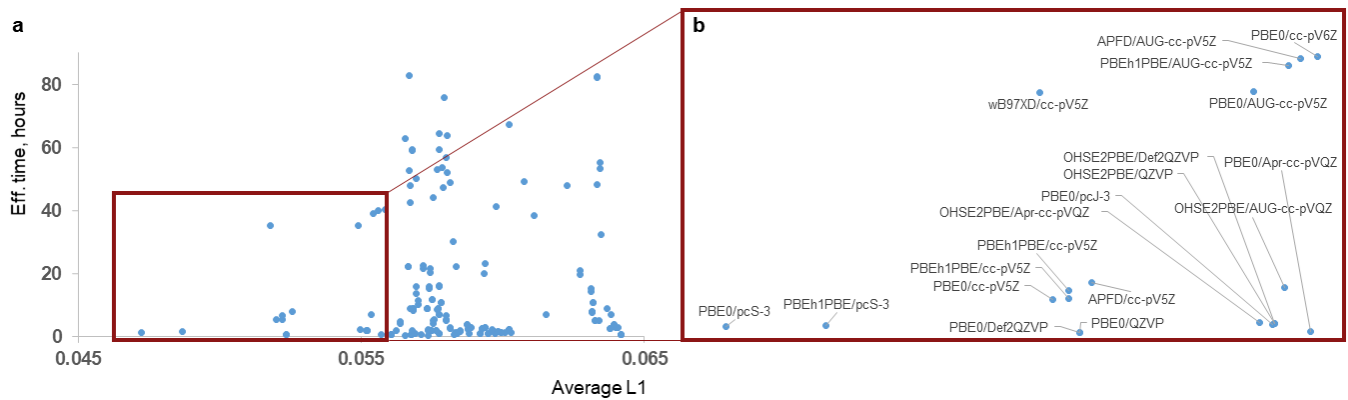

**Fig. 1**. Comparison of accuracy levels (estimated by L1 averaged over first six molecules from the QM9 database) and computation cost (estimated by 'effective time', based on wallclock time for running computations for QM9 entries 8 000, 16 000 and 32 000) for various combinations of a DFT functional and a basis set. Attractive options are PBE0/pcS-3 (which is #4 on the list in Table 1), PBE1hPBE/pcS-3, PBE0/Def2-QZVP, and possibly ωB97X-D/cc-pV5Z. (a) All functional/basis set combinations with average L1 < 0.065 shown; (b) combinations with L1 < 0.056, with labels.

Based on the provided data, we chose the combination of PBE0 functional and pcS-3 basis set to compute DFT electron densities for molecules in the QM9 dataset. This combination has a reasonable computational cost (it took us a couple months to generate the data for the QM9 dataset). We also expect that it has a relatively good accuracy in predicting the electron densities, as follows from the results in the literature and our results on the first six QM9 molecules (Table 1).

We do not claim here that a certain DFT functional or a basis sets is "the best". Different research goals (and hence, metrics of performance, not only the L1 measure used in this work) and different availability of computational resources may make different choices of DFT functionals and basis sets optimal under different circumstances. Besides that, such a claim would incite hatred against us from $N_L - 1$ laboratories, where $N_L$ is the total number of laboratories that have ever published a DFT functional.

To overcome the accuracy of PBE0/pcS-3, CCSD level of theory with VQZ-level or pc-3-level basis sets may be required

To compare the performance of *ab initio* quantum chemical methods to the performance of DFT reported above, we have carried out similar computations for the first six molecules in the QM9 dataset with Hartree-Fock (HF), MP2 and CCSD methods, and various basis sets (Table 2). As expected, in large basis sets, the performance of the methods, in terms of average L1 values, is as follows: CCSD < MP2 < HF (from the most precise to the least precise method).

DFT methods from the top of the list in Table 1, such as ωB97X-D3 and PBE0, have slightly lower average L1 values (~0.05) than MP2 with large basis sets (~0.06), though the ratio of L1 for different molecules differ (for methane MP2 works slightly better). However, many popular DFT functionals, such as B3LYP or TPSSh (L1 ~ 0.06-0.10), are outperformed (in terms of L1 measure) by MP2 in large basis sets.

The performance of CCSD with pcS-3- or VQZ-level basis sets (L1 ~ 0.01-0.03) is consistently better than that of DFT, but the use of smaller (VTZ- or pcS-2-level) basis sets cancels out this advantage of CCSD (L1 ~ 0.04-0.06). In some cases, we have not been able to converge CCSD computations with large basis sets for some of these six molecules ('nan' values in Table 2).

Due to high cost of CCSD computations in large basis sets, especially for molecules like the overwhelming majority of molecules in the QM9 database, it is currently impractical to use such methods to generate datasets of electronic densities comparable in size to QM9. As for smaller basis sets (e.g., VTZ-level sets),[23,52] from the viewpoint of L1 metric of the total electron densities, even the use of CCSD without a frozen core approximation do not offer an increase in performance in comparison to much faster and better scaling DFT methods. The same refers to MP2 methods, regardless of the used basis set: in the best case scenario, they can offer only a marginal improvement in L1 values, not worth the increase in the computational cost (MP2 calculations are more expensive than PBE0).

**Table 2.** L1 measures of the difference between 3D electron densities predicted by two main *ab initio* methods (MP2 and CCSD) with various basis sets, and 'exact' (CCSD/cc-pCV5Z) electron densities [eq. (3)]. Data are given for the first six molecules in the QM9 database. Several top basis sets are given for each theory level, in the order of increasing average value of L1; the values for HF in a large basis set are also shown for comparison. Double horizontal lines between CCSD/cc-pVQZ and CCSD/pcJ-2 shows that some basis sets with intermediate average L1 were omitted.

| method | basis set | 1 | 2 | 3 | 4 | 5 | 6 | average |
|---|---|---|---|---|---|---|---|---|
| MP2 | cc-pVQZ | 0.044 | 0.055 | 0.053 | 0.069 | 0.080 | 0.078 | *0.063* |
| MP2 | cc-pCVQZ | 0.045 | 0.055 | 0.054 | 0.071 | 0.083 | 0.079 | *0.064* |
| MP2 | def2-QZVP | 0.044 | 0.061 | 0.064 | 0.069 | 0.079 | 0.081 | *0.066* |
| MP2 | pcS-3 | 0.043 | 0.060 | 0.061 | 0.075 | 0.083 | 0.081 | *0.067* |
| MP2 | aug-cc-pVQZ | 0.047 | 0.064 | 0.068 | 0.071 | 0.079 | 0.083 | *0.069* |
| MP2 | pc-3 | 0.044 | 0.062 | 0.063 | 0.078 | 0.086 | 0.084 | *0.069* |
| MP2 | pcJ-3 | 0.045 | 0.061 | 0.063 | 0.078 | 0.086 | 0.083 | *0.069* |
| MP2 | aug-cc-pCVQZ | 0.048 | 0.065 | 0.069 | 0.073 | 0.082 | 0.084 | *0.070* |
| MP2 | pcseg-3 | 0.046 | 0.063 | 0.063 | 0.078 | 0.085 | 0.087 | *0.070* |
| CCSD | pcJ-3 | 0.012 | 0.016 | 0.017 | 0.015 | 0.015 | 0.018 | *0.015* |
| CCSD | pcS-3 | 0.011 | 0.017 | 0.018 | 0.017 | 0.017 | 0.019 | *0.016* |
| CCSD | pc-3 | 0.013 | 0.018 | 0.019 | nan | 0.020 | 0.023 | *0.019* |
| CCSD | def2-QZVP | 0.015 | 0.020 | 0.021 | 0.017 | 0.017 | 0.022 | *0.019* |
| CCSD | aug-pcS-3 | 0.022 | 0.018 | 0.019 | nan | 0.017 | 0.020 | *0.019* |
| CCSD | def2-QZVPD | nan | 0.020 | 0.021 | 0.017 | 0.017 | 0.022 | *0.020* |
| CCSD | aug-pc-3 | nan | 0.018 | 0.021 | 0.020 | 0.020 | 0.023 | *0.020* |
| CCSD | aug-cc-pCVQZ | nan | 0.022 | 0.025 | nan | nan | 0.026 | *0.024* |
| CCSD | aug-cc-pVQZ | nan | 0.024 | 0.027 | 0.022 | 0.023 | 0.030 | *0.025* |
| CCSD | cc-pCVQZ | 0.017 | 0.040 | 0.037 | 0.020 | 0.020 | 0.029 | *0.027* |
| CCSD | cc-pVQZ | 0.019 | 0.042 | 0.039 | 0.024 | 0.023 | 0.032 | *0.030* |
| CCSD | pcJ-2 | 0.034 | 0.044 | 0.049 | 0.038 | 0.043 | 0.056 | *0.044* |
| CCSD | aug-cc-pCVTZ | 0.043 | 0.051 | 0.058 | 0.052 | 0.057 | 0.072 | *0.055* |
| CCSD | aug-cc-pVTZ | 0.046 | 0.053 | 0.061 | 0.054 | 0.061 | 0.076 | *0.058* |
| HF | cc-pCV5Z | 0.109 | 0.114 | 0.122 | 0.243 | 0.238 | 0.241 | *0.178* |

<u>DNN trained on DFT data can predict differences in electron density stemming both from electron correlation and small basis set artifacts, in addition to energies</u>

We have designed a DNN that successfully predicts DFT electron densities and energies of molecules from the QM9 database. The only input to the DNN is an approximate electron density of a molecule of interest quickly computed with HF in a small basis set (HF/cc-VDZ), further denoted as $\rho_{HF}$; the same computation also produces an energy, further denoted as $E_{HF}$. The goal of the DNN is to predict the PBE0/pcS-3 electron density (further denoted as $\rho$) and energy (further denoted as $E$) of this molecule. To increase the accuracy with which $\rho$ and $E$ are predicted, the immediate output from the DNN are $\Delta\rho$ and $\Delta E$, where $\Delta\rho$ is the difference between $\rho$ and $\rho_{HF}$ [see eq. (4) in Methods], and $\Delta E$ is the difference between $E$, $E_{HF}$, and a simple linear correction based on the stoichiometry of the molecule [eq. (5), Methods]. Three-dimensional functions $\rho$, $\rho_{HF}$, and $\Delta\rho$ are represented on 64 × 64 × 64 cubic grids. The reported DNN includes a block with ten hidden layers and a U-net architecture,[56,57] with multiple 3D convolution, concatenation and rectified linear unit (ReLu) operations in it. After the U-net block, computations forks into two paths, one leading to $\Delta\rho$ and the other to $\Delta E$. The DNN is trained by minimization of a loss function that includes the terms penalizing deviations of $\Delta\rho$ and $\Delta E$ predicted by the DNN from the corresponding ground truth (i.e., computed from DFT) values. The DNN reported here was trained on 38 268 molecules and validated on 9 537 molecules from the QM9 database; in the final publication, we are planning to present the results of training, validation and testing on the whole QM9 database. For more details on the architecture and training the DNN, see Methods.

The performance of the DNN in predicting $\Delta\rho$ was quantified by $L_{\Delta\rho}rel$ strictly defined in Methods, eq. (10). Intuitively, $L_{\Delta\rho}rel$ shows how different the predicted values of $\Delta\rho$ are from the ground truth values, based on L1 measure. The value of $L_{\Delta\rho}rel = 0$ corresponds to a perfect prediction of DFT results by the DNN, and $L_{\Delta\rho}rel = 1$ means that the DNN is as bad as HF/cc-VDZ in predicting electron densities. Over the validation set of 9 537 molecules, $L_{\Delta\rho}rel$ of the reported DNN after training equals 0.128, and hence the DNN predictions are much closer to the DFT results than to the approximate electron densities computed with HF/cc-VDZ. As for the energy predictions, the reported DNN over the same validation set had the mean absolute error [MAE, see eq. (11)] of 1.07 kcal/mol, which is comparable to MAE values of other ML models for the QM9 dataset reported in the literature (~1 kcal/mol).[17,19,20,25-33]

To further illustrate the performance of the DNN, we consider in more detail its predictions for two specific molecules, namely QM9 entries 110118 and 133119. These two molecules contain a diverse set of functional groups, and thereby can compactly illustrate the DNN performance for various classes of organic substances. According to the scheme of splitting of the QM9 database to training, validation and test sets chosen here (see Methods), these two molecules should belong to the validation set. However, both molecules were not included either into the training or validation sets used to build the reported DNN (quantum chemical computations for both molecules were carried out after the DNN had been trained), and hence they can serve as test cases for the reported DNN. We reserve the test set of QM9 entries as defined in Methods for later analysis, and do not touch them now.

QM9 entry 110118 contains alcohol, amine and nitrile functional groups, a strained ring, as well as C–C and C–H bonds outside the ring that can be considered comparable to some extent in terms of electron density to C–C and C–H bonds in saturated aliphatic compounds (Fig. 2a). The input density $\rho_{HF}$ is mainly localized on non-hydrogen atoms (more precisely, in the close vicinities of the corresponding atomic nuclei), where it ranges from $\rho_{HF} \sim 0.8$ to 20.34 (here and below $\rho$ and $\Delta\rho$ are given in Hartree atomic units of inverse volume, that is bohr$^{-3}$) and on covalent chemical bonds, where $\rho_{HF} \sim 0.2$ (Fig. 2b,c). Outside these regions, $\rho_{HF}$ rapidly decays to zero, reaching negligible values well within the employed grid cube (Fig. 2d). Due to the use of tanh layer (see Methods), the input signal is saturated where $\rho_{HF} \gtrsim 0.8$, which ensures that the information on chemical bonding is not dwarfed by the atomic core densities, and artifacts of a discrete representation of the density near the atomic cores are removed.

The overall performance of the DNN in predicting $\Delta\rho$ of this molecule is slightly worse than that for the validation test set: $L_{\Delta\rho}rel$ is 0.142 for QM9 entry 110118, vs. 0.128 for the validation set. Over the grid points, $\Delta\rho$ ranges from –0.0165 to 0.0386, while the difference of electron densities predicted by the DNN $\Delta\rho_{DNN}$ ranges from –0.0172 to 0.0394.

The lowest values of $\Delta\rho$ are observed in the vicinities of the lone electron pairs of the N and O atoms in the secondary amine and hydroxyl groups, respectively [one lone pair in the N atom and two lone pairs in the O atom, labeled as LP-N(H) and LP-O in Fig. 2e]. Interestingly, this effect stems from two sources of comparable magnitudes: One is electron correlation in these lone pairs, and the other is small basis set artifacts in the HF computation (Fig. 3a,b). The DNN successfully captures the net result, with a quantitative agreement between $\Delta\rho$ and $\Delta\rho_{DNN}$ values (Fig. 2e). Relatively low values of $\Delta\rho$ also occur on the lone electron pair of the N atom in the cyano group [labeled as LP-N(C) in Fig. 2f], in a toroidal region between C and N atoms in the same cyano group (T-CN), near the N atom in the amine group (on the side opposite to the lone pair, OLP-N), and in three regions in the plane of the cyclopropane ring lying on the outer sides of the C–C bonds (OR-CC). Regions LP-N(C) and OR-CC are primarily due to electron correlation, OLP-N is primarily a small basis set artifact, and T-CN comes from both sources (Fig. 3a,b). The DNN reproduces the net effect in of all these regions, with a quantitative agreement in most of them (Fig. 2f). Some difference between the predicted and ground truth values is observed in OLP-N and one of three OR-CC regions (shown by thick yellow arrows in Fig. 2f). However, the isosurface drawn for $\Delta\rho_{DNN} = -0.0073$ much better fits the isosurface for $\Delta\rho = -0.008$ in these regions (data not shown), suggesting that the difference between the DNN predictions and ground truth in this case is mainly caused by an error in the scale, but not the local spatial behavior, of predicted $\Delta\rho_{DNN}$ values in these regions. Next, the isosurfaces for $\Delta\rho$ and $\Delta\rho_{DNN} = -0.003$ (Fig. 2g) demonstrate, in addition to the features discussed above, toroidal shapes around all C–H bonds (T-CH), a toroidal shape around the C–C bond in the side chain (T-CC) – but not between C atoms in the cyclopropane ring – and complex shapes around O–H and N–H moieties that look like mergences of toroidal shapes around N–H and O–H bonds and above-mentioned shapes around lone electron pairs on the N and O atoms. In all regions, the agreement between two isosurfaces for $\Delta\rho$ and $\Delta\rho_{DNN} = -0.003$ is as good as it could be with a grid discretization (Fig. 2g). Interestingly, the toroidal shapes of isosurfaces in regions T-CH and T-CC are formed by an interplay of electron correlation and small basis set artifacts. Electron correlation decreases the density between C and H atoms (C and C atoms) both on the C–H axes (C–C axes, respectively) and around them (Fig. 3b), while small basis set artifacts act in the opposite direction on the C–H axes (C–C axes) and are negligible away from the C–H axis (C–C axes) (Fig. 3c). The combination of these two effects primarily changes the electron density in toroidal regions around C–H and C–C bonds.

Fig. 2. DNN predicts the electron density in the molecule with QM9 index 110118 with high accuracy. Note that this molecule was not used for training or validation of the reported DNN. (a) Structural formula of the molecule. (b) The only input to the DNN was an approximate electron density computed with a fast and not-so-accurate method (HF/cc-pVDZ). The electron density is mainly localized in the vicinities of nuclei and covalent chemical bonds (isosurface for $\rho_{HF} = 0.22$ Bohr$^{-3}$ shown in *gray*). (c) Due to the use of tanh transformation of the density (see Methods), the input to the DNN saturates within the shown regions (isosurface for $\rho_{HF} = 0.8$ Bohr$^{-3}$). (d) Regions with non-negligible electron density entirely fit into the cube used for the grid representation of the density (isosurface for $\rho_{HF} = 0.0001$ Bohr$^{-3}$). (e) The DNN successfully predicts that the DFT density is particularly lower than the input (HF/cc-pVDZ) density on the lone electron pairs of N and O atoms in hydroxyl and secondary amine groups (isosurfaces for $\Delta\rho = –0.014$ Bohr$^{-3}$; ground truth, *blue*, DNN prediction, *red*). (f) The DFT density is also somewhat lower than the input density near the cyano group and in the cyclopropane ring plane (isosurfaces for $\Delta\rho = –0.008$ Bohr$^{-3}$), which is also successfully captured by the DNN, with the exception of the regions shown by *yellow arrows*. (g) Isosurfaces showing where the DFT density is slightly lower than the input density have complex shapes and include toroidal formations around C–C and C–H bonds (isosurfaces for $\Delta\rho = –0.003$ Bohr$^{-3}$), all of which are accurately predicted by the DNN. (h) The DFT density is slightly higher than the input density mainly in the vicinities of the atomic cores and covalent bonds (isosurfaces for $\Delta\rho = +0.003$ Bohr$^{-3}$). The only region with a noticeable difference between the DNN results and the ground truth is shown by *yellow arrows*. (i) The difference of the DFT and input densities is higher near the nuclei (including hydrogen) and the whole cyano group than near the other covalent bonds (isosurfaces for $\Delta\rho = +0.008$ Bohr$^{-3}$). Errors of the DNN near two H atoms are shown by *yellow arrows*. (j) The largest positive difference between the DFT and input electron densities is observed near the nuclei of the O and N atoms (isosurfaces for $\Delta\rho = +0.023$ Bohr$^{-3}$), which is quantitatively correctly predicted by the DNN. (k) The proposed approach is much faster than DFT. Wallclock times to run HF/cc-VDZ computation (HF), to build and coarse-grain the cube file (cube), and to use the DNN for the prediction (DNN), as well as the sum of these three (total), vs. wallclock time to run PBE0/pcS-3 (DFT) computations for this molecule. See the main text for explanations of labels of the features in panels (e-j), discussions of errors of the DNN, and contributions to $\Delta\rho$ from electron correlation and small basis set artifacts. ▶

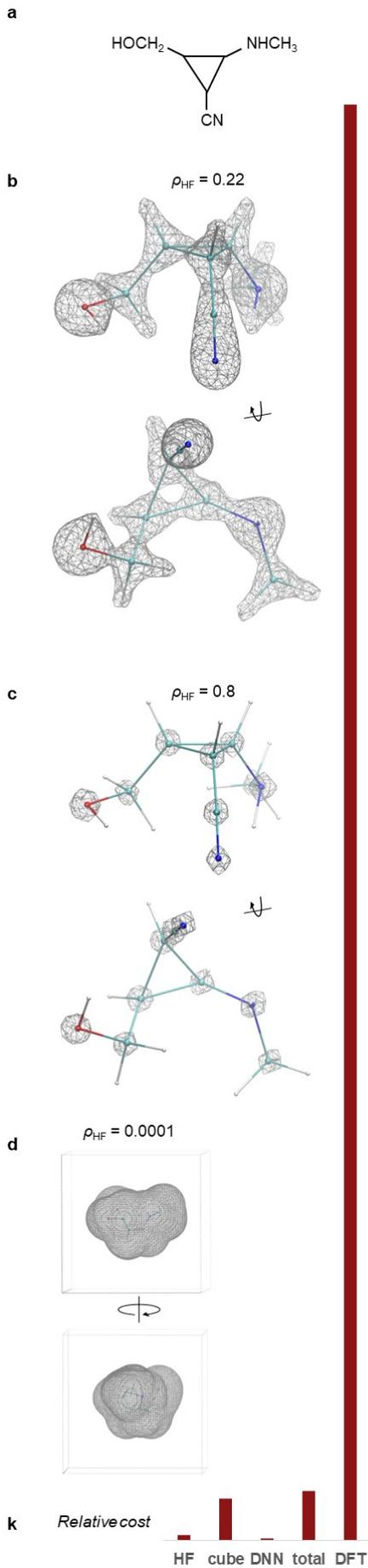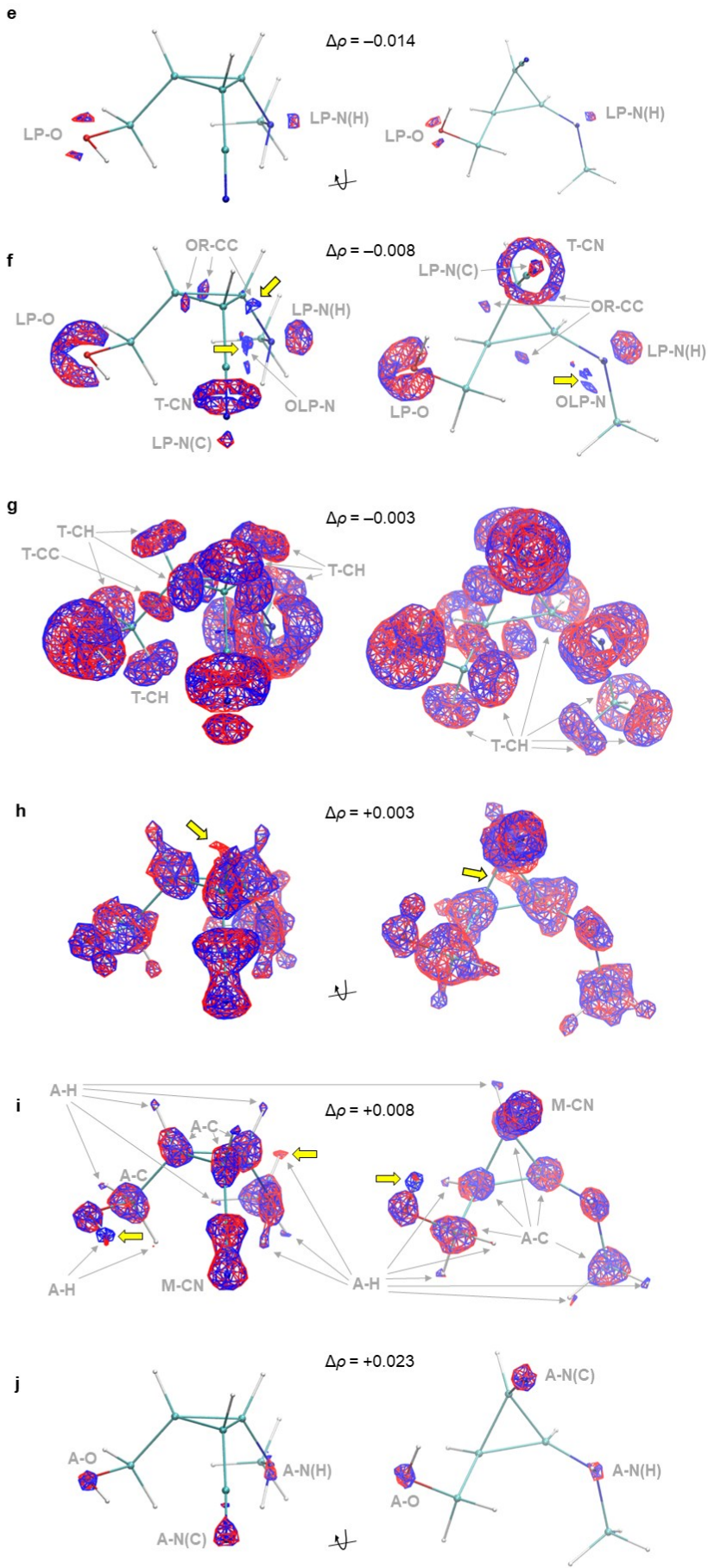

On the opposite end of the $\Delta\rho$ range, the highest values of $\Delta\rho$ are observed around the O and both N nuclei in this molecule [regions labeled A-O, A-N(H), A-N(C), Fig. 2j], and these phenomena come from electron correlation (Fig. 3d). Therefore, the electron density 'leaking' from the lone pairs and OR-CC regions due to electron correlation, as discussed above, ends up mainly in the cores of the N and O atoms (as predicted by DFT). We also notice that the isosurface on the N atom in the cyano group, unlike the N atom in the amine group, is elongated towards the neighboring C atom, and an additional smaller region with a large positive value of $\Delta\rho$ is observed on the C≡N bond, closer to the C atom, both of which being small basis set artifacts (Fig. 3c,d). All these features in $\Delta\rho$ are quantitatively correctly predicted by the DNN (Fig. 2j). Smaller positive values of $\Delta\rho$ are observed on the C and H atoms (regions A-C and A-H, respectively), while the isosurfaces on the C and N atoms in the cyano group at this level of $\Delta\rho$ merge into a single surface (M-CN, Fig. 2i). A-C regions stem chiefly from electron correlation, while A-H and M-CN regions are small basis set artifacts (Fig. 3c). The agreement between the ground truth and DNN predictions is quantitatively correct in most regions, except that the DNN overestimates $\Delta\rho$ on one H atom in the methyl group and underestimates $\Delta\rho$ on the H atom in the hydroxyl (*yellow arrows*, Fig. 2i). In both cases, the differences in the isosurfaces vanish after a small change in $\Delta\rho_{DNN}$ for which the DNN isosurface is drawn (from +0.008 as shown in Fig. 2i to +0.006 for H in the OH moiety or +0.010 for H in $CH_3$, data not shown). Finally, on the isosurfaces built for $\Delta\rho$ and $\Delta\rho_{DNN} = +0.003$, fragments of the surface around most of the H atoms and the corresponding neighboring heavy atoms merge together (Fig. 2h), forming shapes that resemble, but do not coincide with, the HF total electron density profiles (Fig. 2b,h). The agreement between the ground truth and DNN data is again remarkable. The only noticeable difference at this level of $\Delta\rho$ is a protrusion on the isosurface for the DNN data on one side of the cyclopropane ring (*yellow arrow*, Fig. 2h). This shape, however, is matched by a similar one on the ground truth isosurface drawn for a slightly different value of $\Delta\rho = +0.0028$ (data not shown). Therefore, as in the previous cases, the difference between the DNN predictions and ground truth is mainly due to a minor error in the scale of the predicted values of the density differences, but not their qualitative behavior as a function of spatial coordinates.

As for the energy, the correction to the linear estimate for the energy, as predicted by the DNN, is +11.7 millihartree (mh), while the ground truth value computed from the DFT result is +10.4 mh (Table 3). Therefore, the error in the total energy of QM9 entry 110118 computed from the DNN output equals +1.4 mh, or +0.9 kcal/mol. For comparison, MAE of predicted energies over the validation set is 1.07 kcal/mol.

**Table 3**. Energies predicted by the reported DNN differ from the ground truth (DFT) values by ~ 0.002 hartree (~1 kcal/mol) for both molecules considered in detail in the main text. All energies are in hartree.

| QM9 index | $E_{HF}$ | Linear correction | DNN correction | Total predicted $E$ | $E$ | Error in predicted $E$ |
|---|---|---|---|---|---|---|
| 110118 | -416.7604 | -2.2707 | 0.0117 | -419.0194 | -419.0208 | +0.0014 |
| 133119 | -459.4539 | -2.2847 | -0.0005 | -461.7390 | -461.7371 | -0.0019 |

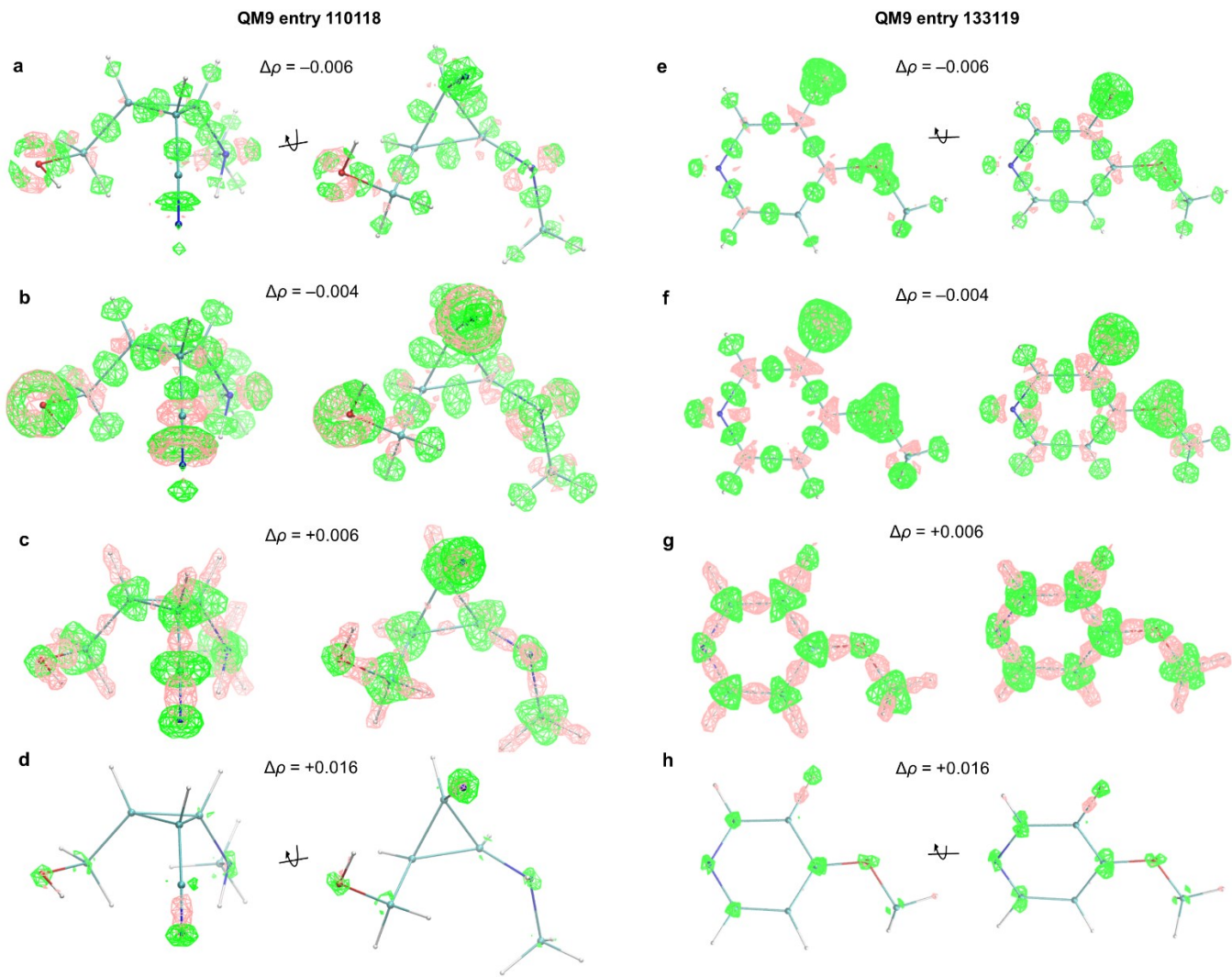

**Fig. 3**. Electron correlation and small basis set artifacts make nontrivial contributions to the difference between the DFT and input (HF/cc-pVDZ) electron densities. The former effect is shown by isosurfaces for the difference between the PBE0/pcS-3 and HF/pcS-3 electron densities (*green*), and the latter by isosurfaces for the difference between the HF/pcS-3 and HF/cc-pVDZ electron densities (*pink*). Both molecules analyzed in the main text, QM9 entry 110118 (a-d) and 133119 (e-h), are shown. (a) In the first molecule, the negative values of $\Delta\rho$ in regions OR-CC and LP-N(C) come mainly from electron correlation, in region OLP-N mainly from small basis set artifacts, and in regions LP-O and LP-N(H) from both sources. (b) In region T-CN, both effects lower the electron density. (c) Electron correlation increases electron density in regions A-C, A-N(H), A-N(C) and A-O, while small basis set artifacts do so in regions A-H, A-N(H), A-N(C) and A-O. Interestingly, electron correlation lowers the electron density between atoms linked with covalent bonds (a,b), but this effect is canceled by an opposite effect of small basis set artifacts (c). Regions T-CH and T-CC emerge as toroidal shapes around C–H and C–C bonds due to cancelations of the effects of electron correlation and small basis set in the middle between the corresponding atoms, but not around these bonds. (d) Particularly strong contributions to the electron density difference come from electron correlation in regions A-N(H), A-N(C) and A-O, as well as basis set artifacts in region M-CN. (e) In molecule with QM9 index 133119, both electron correlation and small basis set artifacts lower the density in regions LP-F, LP-O and LP-N, though electron correlation plays the major role. (f) Electron correlation also lowers the density between the atoms in the aromatic ring, and in wide vicinities of the C–H bonds. (g) These effects are partially canceled out by artifacts of the small basis set on the axes between the corresponding atoms, but not away from such axes, leading to the existence of regions T-CH, P-aCC and P-aCN in the isosurfaces for the total difference of the densities. Additionally, the four-legged merged shape of P-aCN is due to the small basis set effect on the inner side of the N atom, as shown in (f). Electron correlation causes lower electron densities on the F, O and N atoms (regions A-F, A-O and A-N), while the small basis set artifact in the middle of the C–F bond brings about region M-CF in Fig. 4i.

Next, QM9 entry 133119 contains a heterocyclic aromatic ring, an ether functional group, a fluorine atom involved in a C–F bond, as well as C–H bonds in the terminal methyl moiety comparable to some extent in terms of electron density to C–H bonds in saturated aliphatic compounds (Fig. 4a). As in the previous case, the input density $\rho_{HF}$ is mainly localized on non-hydrogen atoms, where it ranges from $\rho_{HF} \sim 0.8$ to 17.99 (the tanh transformation of the input saturates in these regions), and on covalent chemical bonds, where $\rho_{HF} \sim 0.2$ (Fig. 4b,c). Outside these regions, $\rho_{HF}$ rapidly decays to zero, reaching negligible values well within the used grid cube (Fig. 4d).

The overall performance of the DNN in predicting $\Delta\rho$ of this molecule is slightly worse than that for the validation test set: $L_{\Delta\rho}rel$ is 0.158 for QM9 entry 133119, vs. 0.128 for the validation set. Over the grid points, $\Delta\rho$ ranges from –0.0160 to 0.0426, while the difference of electron densities predicted by the DNN $\Delta\rho_{DNN}$ ranges from –0.0152 to 0.0375.

The lowest values of $\Delta\rho$ are observed in the vicinities of the lone electron pairs of the F, O and N atoms (one lone pair in the N atom, two lone pairs in the O atom and three lone pairs in the F atom, labeled as LP-N, LP-O and LP-F in Fig. 4e, respectively). As in the previous case, both electron correlation and small basis set artifacts in the HF/cc-pVDZ density contribute to this effect, though electron correlation plays the major role, especially for the F atom (Fig. 3e,f). The DNN reaches a quantitative agreement between $\Delta\rho$ and $\Delta\rho_{DNN}$ values on the O and N atoms, but not on the F atom, where the DNN significantly underestimates the density difference (Fig. 4e, *yellow arrows*). However, at less negative values of $\Delta\rho$ the DNN predictions get closer to the ground truth, reaching a quantitative agreement by $\Delta\rho \sim -0.008$ (Fig. 4f). At $\Delta\rho$ and $\Delta\rho_{DNN} = -0.003$ (Fig. 4g) the isosurfaces include, besides the above-mentioned features, toroidal shapes around all C–H bonds (T-CH) and pairs of formations on both sides of the aromatic ring where π-bonds between C atoms are localized (P-aCC); similar shapes for two π-bonds between C and N atoms from both sides of the ring merge together into a four-legged-stool shape (P-aCN). In all regions, the agreement between two isosurfaces for $\Delta\rho$ and $\Delta\rho_{DNN} = -0.003$ is as good as it could be with a grid discretization (Fig. 4g). Similar to the toroidal shape of T-CH, the specific form of P-aCC regions comes from an interplay of electron correlation and small basis set artifacts. Electron correlation decreases the electron density around both σ- and π-bonds linking heavy atoms in the aromatic ring (Fig. 3e,f), while small basis set artifacts act in the opposite direction on the σ-bonds and are much less pronounced on the π-bonds (Fig. 3g). As for the specific shape of region P-aCN, it is due to small basis set artifacts lowering the density difference near the N atom on the side opposite from the lone pair (Fig. 3f), which joins four regions on two sides of the ring and near different C–N bonds ('four legs of a stool') into a single connected region.

The highest values of $\Delta\rho$ are observed around the F, O and N nuclei (regions labeled A-F, A-O and A-N, Fig. 4j) due to electron correlation (Fig. 3h). All these features in $\Delta\rho$ are quantitatively correctly predicted by the DNN (Fig. 4j). Smaller positive values of $\Delta\rho$ are observed on the C and H atoms (regions A-C and A-H, respectively), while the isosurfaces on the C and F atoms at this level of $\Delta\rho$ are joined by a bridge (M-CF, Fig. 4i). A-C regions stem chiefly from electron correlation, while A-H and M-CF regions are small basis set artifacts (Fig. 3g). The agreement between the DNN predictions and ground truth is quantitative. Finally, on the isosurfaces built for $\Delta\rho$ and $\Delta\rho_{DNN} = +0.003$, fragments of the surface around most of the H atoms and the corresponding neighboring heavy atoms merge together (Fig. 4h), forming shapes that resemble, but do not coincide with, the HF total electron density profiles (Fig. 4b,h). Also, the DFT density is higher in the middle of the aromatic ring (region M-ar). The agreement between the ground truth and DNN predictions is again remarkable. The only noticeable difference at this level of $\Delta\rho$ is a connection [labeled M-ar-C(F)] between region M-ar and region A-C on the C atom linked to the F atom, which is present in the isosurface built from the DNN prediction, but not in the ground truth isosurface (*yellow arrow*, Fig. 4h). However, the same connection of M-ar and A-C regions appears on the isosurface drawn for a slightly lower value $\Delta\rho = +0.0026$ (data not shown), implying that the error is mainly in the scale of the predicted values of $\Delta\rho$, but not their local spatial behavior.

**Fig. 4**. DNN predicts the electron density in the molecule with QM9 index 133119 with high accuracy. This molecule was not used for training or validating the DNN. (a) Structural formula of the molecule. (b) The input density is mainly localized in the vicinities of nuclei and chemical bonds (isosurface for $\rho_{HF} = 0.22$ Bohr$^{-3}$). (c) The tanh transformation of the density saturates within the shown regions (isosurface for $\rho_{HF} = 0.8$ Bohr$^{-3}$). (d) Regions with non-negligible electron density entirely fit into the grid cube (isosurface for $\rho_{HF} = 0.0001$ Bohr$^{-3}$). (e) The DFT density is particularly lower than the input density on the lone electron pairs of the F, O and N atoms. The DNN successfully predicts these phenomena for the O and N atoms, but not for the F atom (isosurfaces for $\Delta\rho = -0.012$ Bohr$^{-3}$; ground truth, *blue*, DNN prediction, *red*; discrepancy, *yellow arrows*). (f) The isosurfaces drawn for somewhat higher $\Delta\rho$ and $\Delta\rho_{DNN}$ values are in a good agreement, including the toroidal shape around the F atom (isosurfaces for $\Delta\rho = -0.008$ Bohr$^{-3}$). (g) Isosurfaces showing where the DFT density is slightly lower than the input density have complex shapes, and include toroidal formations around C–H bonds and pairs of clusters on the opposite sides of the aromatic ring between C and C or C and N atoms (isosurfaces for $\Delta\rho = -0.003$ Bohr$^{-3}$), all of which are accurately predicted by the DNN. (h) The DNN correctly predicts that the DFT density is slightly higher than the input density mainly in the vicinities of atomic cores and covalent bonds, as well as in the center of the aromatic ring (isosurfaces for $\Delta\rho = +0.003$ Bohr$^{-3}$). However, the difference of densities between the center of the ring and the C atom involved in the C–F bond is overestimated by the DNN (*yellow arrows*). (i) The difference of the DFT and input densities is higher near the atomic nuclei (including hydrogens) and along the C–F bond (isosurfaces for $\Delta\rho = +0.008$ Bohr$^{-3}$), which is correctly predicted by the DNN. (j) The largest positive difference between the DFT and input electron densities occurs near the nuclei of the F, O and N atoms (isosurfaces for $\Delta\rho = +0.023$ Bohr$^{-3}$), which is correctly predicted by the DNN, though the DNN slightly overestimates the difference on the F atom. (k) The proposed approach is much faster than DFT. Wallclock times to run HF/cc-VDZ computation (HF), to build and coarse-grain the cube file (cube), and to use the DNN for the prediction (DNN), as well as the sum of these three (total), vs. wallclock time to run PBE0/pcS-3 (DFT) computations for this molecule. See the main text for explanations of labels of the features in panels (e-j), discussions of errors of the DNN, and contributions to $\Delta\rho$ from electron correlation and small basis set artifacts. ▶

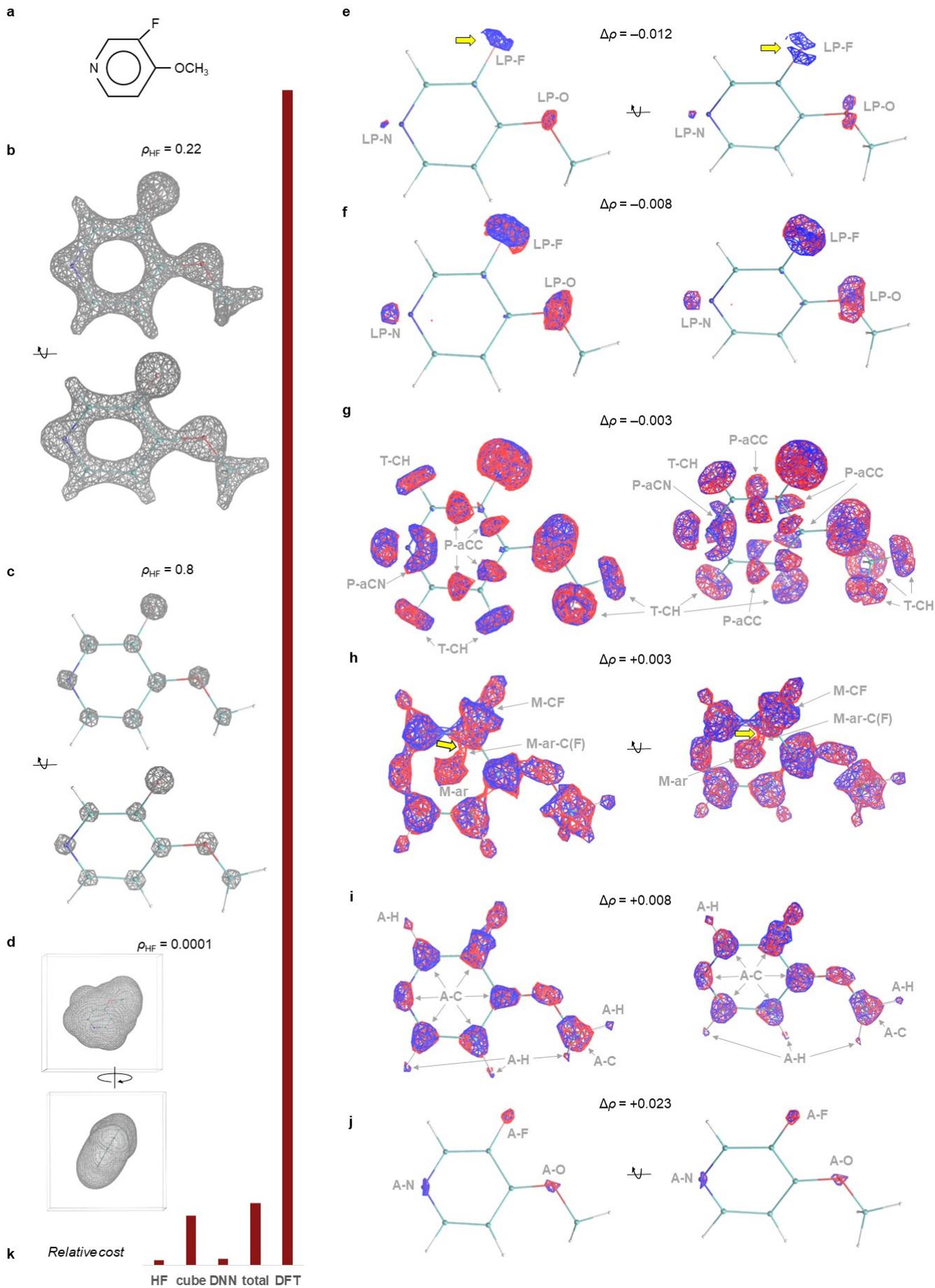

The correction to the linear estimate for the energy predicted by the DNN is –0.4 mh, while the ground truth value is +1.5 mh (Table 3). Therefore, the error in the energy of QM9 entry 133119 predicted with the reported DNN equals –1.9 mh, or –1.2 kcal/mol. For comparison, MAE of predicted energies over the validation set is 1.07 kcal/mol.

The proposed DNN outperforms DFT both in terms of computation speed and scaling with size of molecules

DNN computations run much faster on the same hardware (1 GPU and 2 CPUs, see Methods for details) than DFT (PBE0/pcS-3) computations for every molecule in the QM9 dataset (Fig. 5a). For the overwhelming majority of the database entries (with indices ~100 and higher) the DNN is at least two orders of magnitude faster than DFT. HF computations in the small basis set, required to generate the DNN input, are also much faster than DFT computations for the same molecules. The bottleneck in the current version of the computational pipeline turns out to be the step of generation of input cube files from HF fchk files. To generate cube files, we used cubegen utility from Gaussian, which is currently implemented only on CPUs. We expect that this stage can be significantly speeded up in the future with the use of GPUs. Even in the current setup, the DNN computations, together with HF computations and generation of the input cube files, are faster than DFT (PBE0/pcS-3) computations for all expect the first four molecules in the QM9 database. Closer to the end of the database, the gain in the speed (measured by wallclock time) reaches a factor of ~30. (Only relative wallclock times are shown in this work; absolute values of wallclock times for computations performed in Gaussian are not disclosed according to the Gaussian licensing agreement.)

Wallclock time to run the reported DNN stays virtually the same over the whole QM9 dataset due to the architecture of the reported DNN (Fig. 5a, wallclock time vs. QM9 index; Fig. 5b, wallclock time vs. number of electrons in the molecule). Wallclock time of HF computations increase with the number of elections $N$ as $\sim N^{1.62}$, while PBE0/pcS-3 wallclock time scales less favorably, as $\sim N^{3.26}$. Due to this difference in scaling of the DNN, HF and DFT cost with the size of the molecule, the relative efficiency of the approach proposed here increases with the size of a modeled molecule. The ratio of the wallclock time for HF to that for DFT is ~10% for the first several molecules in the database (entries 1-6), falls to ~1% by entry ~100, and falls even further to ~0.3% closer to the end of the dataset. The wallclock time for the DNN relative to the wallclock time for DFT falls from ~30% for the first several molecules to ~1% by entry ~100, and down to ~0.1-0.3% by the end of the database. As for the relative price of the HF and DNN parts, the DNN stage dominates for smaller systems with up to ~30 electrons. For larger systems, HF becomes consistently more expensive (pronouncedly from ~60 electrons, Fig. 5b), but the gap between the HF and DFT costs in this limit increases. We do not consider here the scaling of wallclock time for the cube file generation, because, as stated earlier, we expect that this part of computations can be significantly accelerated in comparison to the current version.

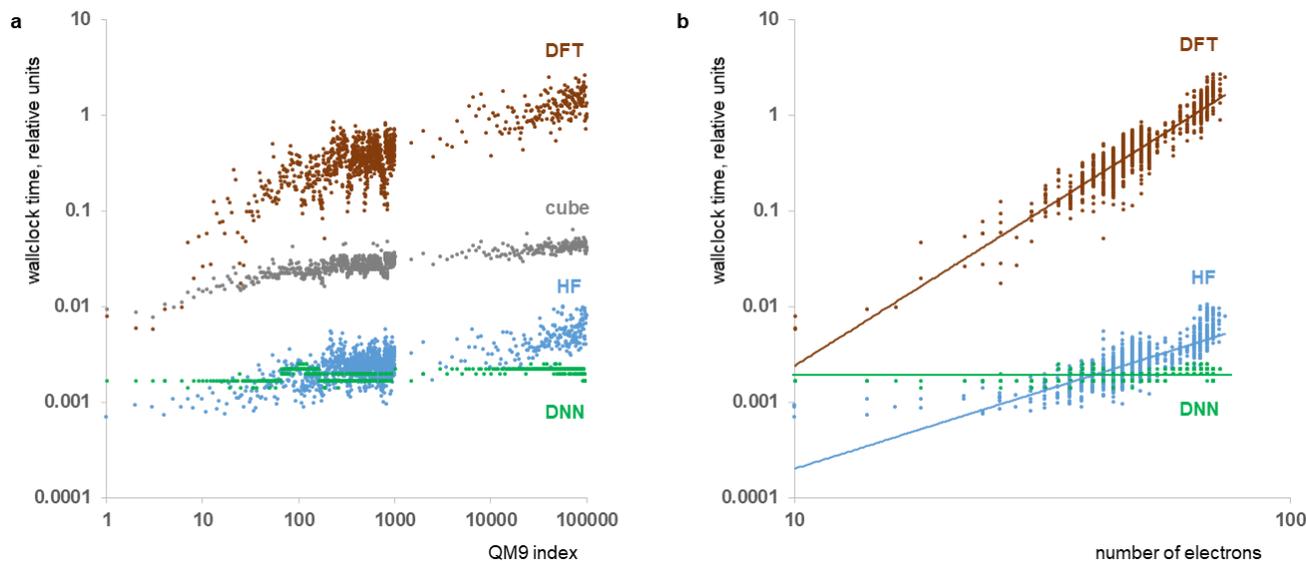

**Fig. 5**. With the approach proposed in this work, electron densities and energies of molecules in the QM9 dataset can be computed much faster than with DFT. (a) Wallclock time to run HF/cc-VDZ computations (HF), to build and coarse-grain the corresponding cube files (cube), and to use the DNN (DNN), vs. wallclock time to run PBE0/pcS-3 (DFT) computations for the same molecules, for various QM9 indices (up to 100,000). Datapoints are given for all molecules with indices up to 1000, and for every 500th molecule with indices from 1,000 to 100,000, hence the difference in the density of points on the plot after QM9 index 1000. (b) HF, DNN and DFT wallclock times as functions of the number of electrons in a molecule. Note log-log scales on both panels. Absolute values of wallclock times for HF and DFT computations performed in Gaussian are not disclosed according to the Gaussian licensing agreement.

For two molecules discussed in detail above, QM9 entries 110118 and 133119, the proposed approach to computing electron densities and energies is faster than DFT computations by factors of 29 and 19, respectively (Fig. 2k and 4k). Durations of HF computations are 0.4% and 0.5%, respectively, of the durations of DFT computations. For the stage of generation of cube files, these ratios are 2.9% and 4.3%, and for the DNN prediction stage itself, these ratios are as low as 0.2% and 0.6%, respectively.

**Discussion**

In the previous literature, ML has been widely used to approximate the dependence of energy on molecular geometry (more specifically, on various descriptors that can be easily computed from the geometry, such as lists of distances to the closest atoms for each atom in the molecule, or more complicated non-quantum-mechanical descriptors).[13,16-21,23,25-28,30-37] The resulting models demonstrated remarkable performance on various test sets of molecules. For example, a number of architecturally diverse models were reported to reach accuracy levels on the order of 1 kcal/mol for organic molecules in the QM9 database (Table 4).[17,19,20,25-33] However, such models might still create an impression of being 'black boxes' that only perform sophisticated 'curve-fitting', rather than learn the real physics of modeled phenomena. From the practical viewpoint, some of these models may turn out to be less transferable to new classes of molecules, physical processes or chemical reactions not included into training and testing datasets. From the viewpoint of theory, it is extremely interesting to understand why such models demonstrate such a good performance, how much 'understanding' of physics such models actually reach, and how this understanding, if any, is encoded in them.

In the present work, to go beyond energy fitting, we build a DNN that also predicts three-dimensional electron densities in molecules. In the reported DNN, computations fork into two paths – one for the energy and the other for the electron density calculations – much closer to the end of the network, thereby pushing the model to learn first some general high-level characteristics, and only after than to employ them to solve the specific tasks of computing the energy or electron density of a given molecule. By including the electron densities into the loss function to be minimized during DNN training, we expect to additionally regularize the model and make it more resistant to overfitting to the specific task of predicting energies. Though some ML models in the literature were trained to predict properties dependent on electron densities, such as partial charges, dipole moments, etc.,[24-26,31,41,52,58,59] and therefore may also be regularized by these additional predictions, the use of electron densities may work better due to a larger number of degrees of freedom in them. Besides that, in order to ensure regularization, predictions of such properties and energies should be done from a shared pool of highly processed features, and will not happen if such predictions are performed by independent neural networks.

Our approach worked surprisingly well for predicting electron densities of organic molecules from the QM9 database. First, it turned out that only one input channel, namely an approximate HF electron density computed in a small basis set, is sufficient for a DNN. Initially, we had expected that several input channels would be required (e.g., gradients and hessians of the approximate density by analogy with DFT, and/or a grid representation of the molecular geometry and nuclear charges, and/or a grid representation of the potential created by the nuclei, etc.). This result implies that even such an imprecise quantum chemical method as HF/cc-VDZ – moreover, only the ground state density computed with it – captures all information relevant for a more accurate quantum description of a molecule, and does so in a form digestible by a DNN. Hence, a promising direction in ML in Quantum Chemistry might be to generate features for ML by a simple-and-fast quantum chemical method, and then to use neural networks to predict results of high-level methods in large basis sets. In a similar manner, it was suggested to use HF matrix elements (but not the electron densities or SCF solutions) to predict MP2 and CCSD energies,[60] or to use MP2 amplitudes to predict coupled cluster energies.[61] Such approaches are computationally more expensive at the stage of featurization, in comparison to using descriptors easily computed from the molecular geometry, but latest advances in speeding up HF computations may mitigate this issue. As demonstrated here, a DNN can correct for small basis set artifacts in the input electron density, which allows for a significant speed up, since HF computations with small basis sets are particularly fast.

Second, the reported DNN predicted multiple features in the density differences, even though the input densities did not contain anything similar to such features. For example, toroidal parts of isosurfaces around C–C and C–H bonds have been placed by the DNN in correct positions, perpendicular to the corresponding chemical bonds, and at a distance where the input density is small by absolute value and does not include any toroidal formations. Also, the DNN learned to distinguish all heavy atoms (C, N, O, F), as seen from correct density difference isosurfaces predicted around them (correspond to 0, 1, 2 and 3 lone pairs, respectively), though no information on the chemical nature of these atoms or their nuclear charges has been explicitly passed to the DNN.

Third, in all cases of significant discrepancies between the DNN predictions and DFT results, we found that errors were in the scale of $\Delta\rho_{DNN}$, but not its local spatial behavior. This might be corrected in future work by adding an additional layer to be learned to perform a nonlinear transformation of the output density, or simply by adding more hidden layers to the network to allow for more flexible nonlinear fitting (though this would slow down computations). Also, the reported DNN may be undertrained on F atoms, in comparison to C, N and O atoms, because the fraction of F-containing molecules in the QM9 database is low. In two cases analyzed in Results in detail, the largest error was found to be in the density differences near the lone electron pairs on the F atom. Hence, DNNs may allow for further improvements in the accuracy of predictions of electron densities.

Table 4. Mean average errors (MAE) of some recent ML models on QM9 and similar datasets. [a] This is the only comparison to coupled cluster energies in this table; all other rows are comparisons to DFT energies.

| Reference | Benchmark | Year | MAE, kcal/mol | Comment |
|---|---|---|---|---|
| 17 | QM9 | 2017 | 0.58 | Kernel ridge regression (with HDAD descriptors), trained on ~ 118 000 molecules, tested on the remaining molecules |
| 19,32 | QM9 | 2017 | 0.31 | Continuous-filter convolutional neural network SchNet (with 6 interaction blocks), trained on 110 462 molecules |
| 20 | QM9 | 2017 | 0.84 | Deep tensor neural network (with 3 interaction passes), trained on 100 000 molecules |
| 18 | ANI-1 | 2017 | 0.78 | Deep neural network ANI, trained on ~14 mln. datapoints (equilibrium and non-equilibrium conformations) for ~56 000 molecules from GDB-8 database. For maximal comparability with the other data in this table, MAE computed on minimum energy conformations of 134 random molecules from GDB-10 database is provided (from Table S3 in the cited paper) |
| 25 | QM9 | 2018 | 3.05 | Kernel ridge regression model (variant $12^{NP}3^B$), trained on the first 3 993 molecules from QM9, MAE computed over the whole QM9 |
| 26 | QM9 | 2018 | 0.30 | Kernel ridge regression-based model, trained on 20 000 molecules and tested on 2 000 molecules |
| 27 | QM9 | 2018 | 1.83 | High-dimensional neural network potentials HDNNPs (with wACSF descriptors), trained (and cross-validated) on 10 000 molecules, tested on ~ 123 000 molecules |
| 28 | QM9 | 2018 | 0.41 | Moment tensor model, trained on 50 000 molecules |
| 29 | QM9 | 2018 | 2.64 | Combination of semiempirical density functional tight-binding method with ML of generalized pair-potentials (with 259 bond types), trained on 2 100 molecules (supplemented with their non-equilibrium conformations), and tested on ~ 130 000 molecules |
| 30 | QM9 | 2018 | 0.26 | Hierarchically interacting particle neural network (with 80 atomic features per layer), trained and tested on ~ 131 000 molecules |
| 31 | QM9 | 2018 | 1.5 | Kernel ridge regression (with $F_{2B} + F_{3B}$ features), trained on 5 000 random molecules and tested on 126 722 molecules |
| 33 | QM9 | 2018 | 0.41 | Neural networks with two hidden square unit augmented layers, trained on 100 000 molecules and tested on ~ 31 000 molecules |
| 38 | GDB07to09 | 2018 | 0.80 | Deep neural network ANI-1x, trained on 25% of ANI-1 database with active learning. For maximal comparability with the other data in this table, MAE computed on GDB07to09 benchmark (1500 molecules with 7, 8 or 9 C, N, O atoms) for conformations with energy within 10 kcal/mol of minima are provided (from Table S15 in the cited paper) |
| 40 | CCSD(T)*/CBS | 2018 | 1.46[a] | Deep neural network ANI-1ccx, obtained from ANI-1x by transfer learning on CCSD(T)* energy data. For maximal comparability with the other data in this table, MAE computed on GDB10to13 benchmark (2996 molecules with 10 to 13 C, N, O atoms) for conformations with energy within 100 kcal/mol of minima are provided (from Table 2 in the cited paper). |
| This work | QM9 | 2018 | 1.07 | Deep neural network, trained on 38 268 molecules and validated on 9 537 molecules |

As for predicting energies, the present DNN reached the accuracy level on the order of 1 kcal/mol, which is on par with models reported in the literature (Table 4). We expect that the accuracy might be further improved, as we have not heavily optimized the architecture of DNN for energy predictions in this work.

Due to the chosen architecture of the DNN, the cost of the DNN computations stays nearly constant with the size of a molecule

(Fig. 5b). However, this relationship holds only for molecules that fit into the cube grid we used in this work. In general, for molecules much larger than those in the QM9 database, larger grids would be required, and a new DNN with a different architecture would have to be trained. We speculate that the cost of DNN predictions in this regime will scale in the range between $\sim N \log N$ and $\sim N^3 \log N$, depending on the shape of molecules. The first factor ($N$ to $N^3$) comes from an increase of the required 3D grid size: for compact molecules, roughly linearly with the molecule volume, hence roughly linearly with $N$, but for elongated molecules, roughly cubically with the molecule length, hence cubically with $N$. The second factor, $\log N$, estimates possible increase in the depth of the network that may be required to process larger grids. The scaling could be made more favorable (maybe down to $\sim N \log N$) with more flexible designs of DNNs, enabling them to work with non-cubic inputs of variable size. This would allow for $\sim N$, not $\sim N^3$, scaling of the input size for elongated molecules. However, such an optimization of DNNs might not be a top priority now, because for large molecular systems, which are of more practical interest, not only cube file generation, but even HF computations are currently more expensive than DNN computations (Fig. 5).

Speaking of possible practical applications of DNNs capable of predicting electron densities, first of all, we note that the knowledge of the electron density is sufficient to compute forces acting on all atoms in a molecular system (more precisely, derivatives of energy over coordinates of nuclei), as follows from the Hellmann-Feynman theorem. We expect that these computations will be numerically more accurate than computations based on direct fits for energies as functions of the molecular geometries, because the former approach involves integration, while the latter involves differentiation of approximate functions. Along this way, one may be able to run *ab initio* molecular dynamics simulations accounting for the correlation energy at a low computational cost. Besides calculations of forces, electron densities predicted by DNNs can be used to study inter- and intramolecular interactions, and to compute any quantum mechanical observables that depend only on one-particle electron density, for example, dipole and higher-order electric moments of molecules.

DNNs can offer a conceptually new solution to the problem of polynomial scaling of the cost of most quantum chemical methods with the size of molecules [or even faster-than-polynomial scaling in full configuration interaction (FCI) computations]. This solution is based on the faster-than-exponential growth of the number of molecules with the number of atoms they consist of, or the number of electrons they have. If a dataset of molecules is prepared by including all or most molecules smaller than a certain threshold size, as is the case of QM9 and many other quantum chemical databases, then most molecules from the dataset will have comparable cost of running high-accuracy quantum chemical computations to get the data for training and testing. For example, the QM9 dataset consists of organic molecules that have up to nine heavy atoms (C, N, O, F) and an arbitrary number of H atoms. Rapid growth in the DFT running time is observed with the growth in the number of heavy atoms (~1 min for molecules with one or two heavy atoms, ~3-7 min for three heavy atoms, up to 15 min for four heavy atoms, and ~0.5 to 2 hours for nine heavy atoms under used conditions; see Methods). For *ab initio* computations, this increase is even more pronounced, because *ab initio* methods have less favorable scaling with the problem size than DFT. However, this sharp increase is localized to a tiny part of the QM9 database (one or two heavy atoms: 8 first entries in QM9 database, or 0.006% of all entries; three heavy atoms: 9 entries, or 0.007%; four heavy atoms, 31 entries, or 0.023%). Out of 133 885 molecules in this database, 111 897 (or 84%) have nine heavy atoms, and comparable costs of running DFT computations (~0.5 to 2 hours under used conditions). As soon as the threshold level of a cost of a single computation is overcome (in this work and with our computational resources, this threshold level was on the order of a few hours per molecule), the cost of building a database increases roughly linearly with the number of database entries, regardless of the scaling of a cost of a single computation with the size of a molecule. And could there be anything nicer than linear scaling? This was a rhetorical question, of course.

It is important to emphasize that deep learning does not replace or invalidate the methods of Quantum Chemistry. ML actually reinforces these methods and extends them to a wider realm of practical applicability. Unlike *ab initio* methods, predictions of electron densities and energies with DNN are expected to have much more favorable scaling, as discussed above. However, this favorable scaling can be achieved only on the basis of running high-level quantum chemical computations for molecules in a sufficiently large training database. The use of DNNs therefore separates the problem of increasing the accuracy of quantum chemical methods from the problem of routine applicability of such methods. With ML, it may become not required that an accurate quantum chemical method works fast enough for every new molecule that end users may be interested in. Instead, the focus shifts to generating highly accurate results only for a finite dataset to be used for training, while the efficiency in practical applications is to be achieved via improvements in DNNs to make them faster and more accurate.

In connection to the problem of training datasets, a question that still remains open is how the performance of various quantum chemical methods, in combinations with various basis sets, relates to each other. This question is very important from the practical viewpoint, because it determines the strategy of generating datasets for ML. Which theory level and basis set to use, given limited computational resources? In this work, we publish the database of DFT (PBE0/pcS-3) solutions for the QM9 dataset. However, these wavefunctions (or electron densities computed from them) are still far from the exact ones, as follows from Tables 1 and 2. What will be the next level of accuracy for QM9 or other quantum mechanical databases of a similar size? Our results imply that one has to do CCSD-level computations with at least VQZ-level basis sets to significantly improve the accuracy. Our data on the relative performance of different methods and basis sets are limited by the used measure L1 and the set of six molecules for which computations were done, and much further work in this direction is required.

Other possible directions for further research include the following. Databases of electron densities for a large number of inorganic molecules, organic molecules with more elements than those in QM9, ions, noncovalently bound molecular complexes and other molecular systems not included in the QM9 database should be built and used for training. New architectures of DNNs, allowing for piecewise scanning of input and prediction of output, could be developed to remove a restriction on the size of a modeled system that exists in DNNs with the reported architecture. The accuracy of DNN predictions and/or their transferability to other classes of molecules might be improved if more input channels are used, for example, gradients (and possibly hessians) of approximate density functions, or explicit information on charges and positions of nuclei in the molecule. Also, the present approach needs to be extended to modeling geometries of molecules far from equilibrium, and processes of bond formation and breaking, which may require abandoning the use of a single HF computation as the DNN input. Finally, it may be possible to improve practical efficiency and simplify physical interpretation of DNNs by using other 3D representations of electron densities, such as non-cubic grids used for numerical integration in DFT (with clusters of integration points centered on nuclei and arranged by radial shells and angular directions). However, it is not evident how to efficiently perform three-dimensional convolution on such data, so this representation may require a radical revision of architectures of used neural networks.

A fundamentally different approach to the use of ML in computations of electron densities and energies of molecules is being actively investigated in the literature. It is based on the idea that the Hohenberg–Kohn theorems[1] state the existence of the energy functional of the electron density, the exact analytical expression of which is not known, but this functional (or one of its nontrivial components) could be learned with the use of ML methods.[14-16,23,62-66] A comparative analysis of this approach and the approach reported in this paper goes much beyond our work. However, we would like to mention that the use of learned energy functional in an iterative minimization procedure may be computationally more expensive than a direct computation of the density as suggested in our work. Also, minimization / maximization tasks for DNNs are known to be associated with artifacts, such as "adversarial examples" (slightly perturbed images that are wrongly classified by DNNs, though the original images were classified correctly).[67,68] Optimization procedures with functions or functionals approximated by DNNs in Quantum Chemistry may encounter similar problems, ending up in unphysical solutions ("adversarial electron densities"). On the other hand, if this issue with false minima does not appear or can be circumvented, then ML of the energy functional may lead to methods more transferable to new classes of molecules not used for training, in comparison to a direct prediction of the densities as suggested in this work.

We would like to end this discussion with a repetition that the key element of success in applications of ML in Quantum Chemistry, in our opinion, lies in involving as much physics as possible into ML models. We demonstrate in this work that the use of HF electron densities, even computed in a small basis set, looks promising, perhaps because such representation may be more physical than other descriptors. Training ML models not only on energies, but also on electron densities may serve as a possible strategy to make models more physical, and stimulate learning the physics of modeled phenomena, rather than curve fitting.

**Conclusion**

The reported results show that DNNs may be a promising tool in augmenting Quantum Chemical computations and enabling high-accuracy simulations of large molecular systems at a low computational cost. The present convolutional DNN works surprisingly well with three-dimensional data on electron densities, both as input to and output from DNNs, and does so much faster than DFT.

**Methods**

Quantum chemical computations

Production DFT (**PBE0/pcS-3**) computations for all molecules in the QM9 database were carried out in Gaussian 16, revision A.03 (Gaussian, Inc.). Each computation used one GPU (NVIDIA Kepler GK210, in NVIDIA Tesla K80) and two CPUs (in Intel Xeon CPU E5-2680 v2) on XStream, a Cray CS-Storm GPU compute cluster at Stanford University (http://xstream.stanford.edu). The geometry of all molecules was taken from the QM9 database. No prior changes in the geometry, such as energy minimization, were performed. Computations in Gaussian were run with DFT functional "PBE1PBE" and keywords "NoSymmetry Output=WFX Density=Current Population=Full". The pcS-3 basis set was downloaded from EMSL Basis Set Library (https://bse.pnl.gov).[69,70] Both wfx and chk files were recorded; fchk files were subsequently generated from chk files with formchk utility from Gaussian. Cube files for the total electron density were generated from fchk files with cubegen utility from Gaussian with keyword "FDensity=SCF". The grid had 256 points per each side (cubic 256 × 256 × 256 grid), and the step size of 0.1 bohr in each direction. Each molecule was positioned in the middle of a cube (i.e., shifted in space relative to the position in the original QM9 database); no rotations of the molecules were performed. The resulting 256 × 256 × 256 cube files were coarse-grained to 64 × 64 × 64 cube files as described below.

Production **HF/cc-VDZ** computations for all molecules in the QM9 database were carried out in the same way as PBE0/pcS-3 computations (see above), except that the method keyword was "HF", and the cc-VDZ basis set internally implemented in Gaussian was used.

Reference **CCSD/cc-pCV5Z** computations for the first six molecules from the QM9 database were carried out in Gaussian (as above). Each computation used one CPU (in Intel Xeon E5-4640, E5-4650v4 or E5-2697Av4) and up to 400 GB of memory on Sherlock, a high-performance computing cluster at Stanford University (https://www.sherlock.stanford.edu). To make the full use of the symmetry of these small molecules, Z-matrices accounting for their high symmetry were manually generated. The values of the bond lengths, angles and dihedral angles were computed from the corresponding Cartesian coordinates in the QM9 database; in the cases of small differences between such values computed from different subsets of atoms related by symmetry operations, arithmetic averages of the values were used. No other changes in the geometry, such as energy minimization, were performed. The basis set was taken from EMSL Basis Set Library (as above). Keywords "Output=WFX Density=Current Population=Full" were used. Both wfx and chk files were recorded; fchk files were subsequently generated from chk files with formchk utility from Gaussian. Cube files for the total electron density were generated from fchk files with cubegen utility from Gaussian with keyword "FDensity=CC". The grid had 161 points per each side (cubic 161 × 161 × 161 grid), and the step size of 0.1 bohr in each direction. No shifts or rotations of the molecules in the cube files were performed (due to the use of Z-matrices rather than coordinates from the QM9 database for these computations).

Various **HF, MP2, CCSD and DFT computations with various basis sets** for the purpose of generating data shown in Tables 1 and 2 were carried out in Gaussian (as above) or Q-Chem, version 5.1.0 (Q-Chem, Inc.). Computations were run on XStream (as above) with the use of one CPU and one GPU (as above), or on Sherlock (as above) with the use of one CPU (as above). Z-matrices, the same as in CCSD/cc-pCV5Z computations, were used. Basis sets internally implemented in Gaussian or Q-Chem were used. For computations in Gaussian, chk files were saved, converted to fchk files, and then cube files (161 × 161 × 161 grid, step size of 0.1 bohr) were computed (as above). For computations in Q-Chem, cube files of the same size were directly generated, with the use of "make_cube_files true" keyword in the Q-Chem input files. No frozen cores were used in any of these computations.

The following DFT functionals were screened (keywords for the methods in the corresponding software are given): APFD, B3LYP, B3PW91, BLYP, HSEH1PBE, M062X, mPW3PBE, OHSE1PBE, OHSE2PBE, PBE1PBE, PBEh1PBE, TPSSh, ωB97X, ωB97XD (in Gaussian), B3LYP, B3PW91, B97-D3, B97M-rV, BLYP, M06-2X, M06-L, PBE, revPBE, revPBE0, TPSS, TPSSh, ωB97M-V, ωB97X, ωB97X-D, ωB97X-D3, ωB97X-V, wM05-D (in Q-Chem). These functionals were selected because they demonstrated high performance in various benchmark studies.[5,6,52]

The list of screened basis sets included the following: Apr-cc-pV5Z, Apr-cc-pV6Z, Apr-cc-pVDZ, Apr-cc-pVQZ, Apr-cc-pVTZ, AUG-cc-pV5Z, AUG-cc-pV6Z, AUG-cc-pVDZ, AUG-cc-pVQZ, AUG-cc-pVTZ, aug-pc-3, aug-pc-4, aug-pcJ-3, aug-pcJ-4, aug-pcS-3, aug-pcS-4, aug-pcseg-3, aug-pcseg-4, CBSB7, cc-pV5Z, cc-pV6Z, cc-pVDZ, cc-pVQZ, cc-pVTZ, CEP-121G, CEP-31G, CEP-4G, D95, D95V, dAug-cc-pV5Z, dAug-cc-pV6Z, dAug-cc-pVDZ, dAug-cc-pVQZ, dAug-cc-pVTZ, Def2QZV, Def2QZVP, Def2QZVPP, Def2SV, Def2SVP, Def2SVPP, Def2TZV, Def2TZVP, Def2TZVPP, DGDZVP, DGDZVP2, DGTZVP, EPR-II, EPR-III, Jul-cc-pV5Z, Jul-cc-pV6Z, Jul-cc-pVDZ, Jul-cc-pVQZ, Jul-cc-pVTZ, Jun-cc-pV5Z, Jun-cc-pV6Z, Jun-cc-pVDZ, Jun-cc-pVQZ, Jun-cc-pVTZ, LanL2DZ, LanL2MB, May-cc-pV5Z, May-cc-pV6Z, May-cc-pVDZ, May-cc-pVQZ, May-cc-pVTZ, MidiX, MTSmall, pc-3, pc-4, pcJ-3, pcJ-4, pcS-3, pcS-4, pcseg-3, pcseg-4, QZVP, SDD, SDDAll, SHC, spAug-cc-pV5Z, spAug-cc-pV6Z, spAug-cc-pVDZ, spAug-cc-pVQZ, spAug-cc-pVTZ, STO-3G, SV, SVP, TApr-cc-pV5Z, TApr-cc-pV6Z, TApr-cc-pVDZ, TApr-cc-pVQZ, TApr-cc-pVTZ, TJul-cc-pV5Z, TJul-cc-pV6Z, TJul-cc-pVDZ, TJul-cc-pVQZ, TJul-cc-pVTZ, TJun-cc-pV5Z, TJun-cc-pV6Z, TJun-cc-pVDZ, TJun-cc-pVQZ, TJun-cc-pVTZ, TMay-cc-pV5Z, TMay-cc-pV6Z, TMay-cc-pVDZ, TMay-cc-pVQZ, TMay-cc-pVTZ, TZV, TZVP, UGBS, UGBS1O, UGBS1P, UGBS1V, UGBS2O, UGBS2P, UGBS2V, UGBS3O, UGBS3P, UGBS3V, 3-21G, 4-31G, 6-21G, 6-311+G, 6-311G, 6-31G (in Gaussian), aug-cc-pCV5Z, aug-cc-pCVDZ, aug-cc-pCVQZ, aug-cc-pCVTZ, aug-cc-pV5Z, aug-cc-pVDZ, aug-cc-pVQZ, aug-cc-pVTZ, aug-pc-1, aug-pc-2, aug-pc-3, aug-pc-4, aug-pcS-0, aug-pcS-1, aug-pcS-2, aug-pcS-3, aug-pcS-4, aug-pcseg-0, aug-pcseg-1, aug-pcseg-2, aug-pcseg-3, aug-pcseg-4, cc-pCV5Z, cc-pCVDZ, cc-pCVQZ, cc-pCVTZ, cc-pV5Z, cc-pVDZ, cc-pVQZ, cc-pVTZ, crenbl, def2-QZVP, def2-QZVPD, def2-QZVPP, def2-QZVPPD, def2-SVP, def2-SVPD, def2-TZVP, def2-TZVPD, def2-TZVPP, def2-TZVPPD, DZ, DZ+, DZ++, G3LARGE, G3MP2LARGE, hwmb, lacvp, lanl2dz, lanl2dz-sv, pc-0, pc-1, pc-2, pc-3, pc-4, pcJ-0, pcJ-1, pcJ-2, pcJ-3, pcJ-4, pcS-0, pcS-1, pcS-2, pcS-3, pcS-4, pcseg-0, pcseg-1, pcseg-2, pcseg-3, pcseg-4, r64G, racc-pVDZ, racc-pVQZ, racc-pVTZ, rcc-pVQZ, rcc-pVTZ, sbkjc, srlc, srsc, STO-2G, STO-3G, STO-6G, SV, TZ, TZ+, TZ++, TZV, UGBS, VDZ,

VTZ, 3-21+G, 3-21G, 4-31G, 6-31+G, 6-311+G, 6-311G, 6-31G (in Q-Chem).

In total, for DFT, we obtained nontrivial results for 1259 combinations of a functional and a basis set in Gaussian, and 1744 combinations in Q-Chem (two lists of combinations partially overlapped). For MP2, we successfully ran computations for at least one molecule out of six with 97 basis sets, and for CCSD, with 86 basis sets. These numbers exclude screened combinations of functionals and basis sets (for DFT) or basis sets (for MP2 and CCSD) for which quantum chemical computations have not converged for various reasons [e.g., insufficient wallclock time (up to 2 days allocated), insufficient memory, not diverged SCF iterations, etc.].

L1 measures reported in Tables 1 and 2 were computed as follows. Cube files for two compared combinations of method and basis set were calculated, either from fchk files (as described) or directly (CCSD and MP2 computations in Q-Chem), with the same size, position and orientation of the grids as for the reference electron densities. After that, differences of two cube files were computed with cubman utility from Gaussian, and sums of absolute values of all elements in each difference cube files were computed with a simple homemade C++ code. The values of the integral in L1 measure were computed as the products of sums of all values multiplied by the grid spacing cubed (essentially, with a 3D generalization of the rectangle rule). Simultaneously, an integral of each electron density over the whole cube was computed to check the accuracy of such integration and the sufficiency of the cube size. We tried different grid spacings and concluded that 0.1 bohr (but not 0.2 bohr) is sufficient to get at least two correct significant figures in the values of L1 measures (data not shown). With this grid spacing, it is sufficient to use a $161 \times 161 \times 161$ grid to fit any of six molecules shown in Table 1 and 2.

To compute the effective time, we run DFT computations for each combination of a functional and basis set for QM9 entries 8 000, 16 000 and 32 000, as described above, and recorded total wallclock time for each computation to complete. Whenever possible, ratios of total times for molecules 16 000 and 8 000 were computed, and a median value of these ratios across all combinations of a functional and basis set was found. Similarly, a median value of the ratio of wallclock times for molecules 32 000 and 8 000 was computed. Finally, for every functional/basis set combination, the wallclock time for molecule 8 000, the wallclock time for 16 000 divided by the median for the 16000/8000 ratio, and the wallclock time for 32 000 divided by the median for the 32000/8000 ratio were computed whenever possible. The effective time for every functional/basis set combination was computed as a geometric average of those of three variables that were available.

DNN architecture and training

As mentioned earlier, the only **input** to the DNN is an approximate electron density of a molecule of interest $\rho_{HF}$. Preparation of an input file is carried out in three steps:

(1) For a given molecule, HF calculations are run in standard quantum mechanical software.
(2) Using the fchk file generated in step (1), a cube file for the total electron density is generated. This cube file contains numerical values of the electron density on $256 \times 256 \times 256$ grid points with a grid spacing of 0.1 bohr (~0.05 Å). The grid spacing is chosen to be the same as in the previous subsections, and the number of grid points in each direction was chosen following a tradition in the field of deep learning to use power-of-two grids, which simplifies architectures of neural networks (a $128 \times 128 \times 128$ grid with a grid spacing of 0.1 bohr is not large enough to contain some of the largest QM9 molecules).
(3) The $256 \times 256 \times 256$ cube file generated in step (2) is coarse grained to a $64 \times 64 \times 64$ cube file by summation of the density values in non-overlapping $4 \times 4 \times 4$ cubes. We perform this transformation to speed up DNN training and make training possible on a single GPU. We checked that this coarse-graining quantitatively preserves the spatial behavior of the electron density. Coarse-graining of a $256 \times 256 \times 256$ cube into a $64 \times 64 \times 64$ cube ensures that the integral of the electron density computed from a sum of the values on all grid points has the right value, and mitigates artifacts of discrete representation of the electron density near nuclei where the gradient of the density is large. A $64 \times 64 \times 64$ cube file with a grid spacing of 0.4 bohr directly generated from the fchk file does not satisfy either of these two conditions, hence the need for separate steps (2) and (3).

The immediate **output** from the DNN are $\Delta\rho$, which is the difference between $\rho$ and $\rho_{HF}$:

$$\Delta\rho(\mathbf{r}) = \rho(\mathbf{r}) - \rho_{HF}(\mathbf{r}), \qquad (4)$$

and $\Delta E$, strictly defined as:

$$\Delta E = E - \left[ E_{HF} + c_0 + \sum_{a \in H,C,N,O,F} c_a n_a \right], \qquad (5)$$

where $n_a$ is the number of atoms of element $a$ in the molecule, and $c_0$ and $c_a$ are empirical coefficients found from the least square fit of the equation

$$E - E_{HF} \approx c_0 + \sum_{a \in H,C,N,O,F} c_a n_a, \qquad (6)$$

over the molecules in the training set ($c_0 = 0.01131$, $c_H = -0.02016$, $c_C = -0.20573$, $c_N = -0.26417$, $c_O = -0.31768$, $c_F = -0.35881$, all values in hartree). In other words, the DNN is used to predict the discrepancies, eq. (5), between the right and left hand sides of eq. (6).

From $\Delta\rho$, the desired values of $\rho$ can be easily computed, because $\rho_{HF}$ is already known. The values of $\Delta\rho$ used for training were computed from PBE0/pcS-3 results similar to $\rho_{HF}$ as described above, and represented after coarse-graining in the form of $64 \times 64 \times 64$ cube files. The output of the DNN follows the same format of $\Delta\rho$ representation. Similarly, the value of $\Delta E$ predicted by the DNN is sufficient to compute the desired value of $E$, because $E_{HF}$ is known from the HF/cc-VDZ computation that we perform anyway to get the input to the DNN, and the linear correction is easy to compute from the molecular formula.

The **architecture** of the DNN is shown in Fig. 6. The only input channel is the HF/cc-VDZ density given on a $64 \times 64 \times 64$ grid. First, this input is processed elementwise with a tanh function [namely, $\tanh(1.28\rho_{HF})$], such that $\rho_{HF}$ in the regions where $\rho_{HF} \gtrsim 0.8$ (typical of atomic cores) saturates to 1, while in other regions (including covalent bonds) the input is only linearly rescaled. This transformation ensures that the information on chemical bonding is not dwarfed by the atomic core densities, and artifacts of a discrete representation of the density near the atomic cores are removed. Next, the information is processed by ten hidden layers with a U-net architecture. This type of architecture, originally proposed to process medical images,[56] and proved efficient in other tasks,[57] but have been used so far only for two-dimensional images, to the best of our knowledge. The first five hidden layers encode step-by-step the input into a very coarse spatial representation, the subsequent five hidden layers decode it back to the original resolution, and there is also a direct flow of information from encoding hidden layers to decoding hidden layers having the same spatial resolution (hence the term 'U-net' for the architecture). Along this path, spatial resolution goes from $64 \times 64 \times 64$ to $2 \times 2 \times 2$ and then back to $64 \times 64 \times 64$, and intermediate representations include up to 256 channels. At the end of the U-net block, a tensor with 64 channels, each of which has a $64 \times 64 \times 64$ spatial resolution, is computed and concatenated with the input (after tanh transformation), yielding a 65-channel tensor. After this concatenation, computations fork into two paths – one for the electron density and the other for the energy calculations. The density is computed by a convolution of the concatenated tensor to 32 channels with the same spatial resolution, rectified linear unit (ReLu) activation, and concatenation of the result to a single channel yielding the predicted $\Delta\rho$ values on a $64 \times 64 \times 64$ grid. The other path of computations includes two subsequent 3D convolution operations, each of which is followed by ReLu activation, such that the first convolution decreases the number of channels to 32, and the second one to 16. Finally, the output value of $\Delta E$ is computed as a linear function of the elements in all 16 channels and on all $64 \times 64 \times 64$ grid points. This forked architecture of the network is designed to push the model to learn first some general high-level features, and only after than to employ these general features to solve the specific tasks of computing the energy or electron density of a given molecule.

The loss function $L$ to be minimized during training was chosen as a linear combination of L1 measures of the performance of the DNN in predicting the electron densities and energies:

$$L = \sum_{i \in \text{training set}} \left\| \Delta\rho_{DNN}(\rho_{HF}(i)) - \Delta\rho(i) \right\|_1 + w_E \sum_{i \in \text{training set}} \left| \Delta E_{DNN}(\rho_{HF}(i)) - \Delta E(i) \right|, \quad (7)$$

where $i$ numerates molecules in the training set, $\Delta\rho(i)$ and $\Delta E(i)$ are the ground truth (i.e., computed from DFT) values of $\Delta\rho$ and $\Delta E$ for the $i$-th molecule, $\rho_{HF}(i)$ is the input (HF/cc-VDZ) electron density for the $i$-th molecule, $\Delta\rho_{DNN}$ and $\Delta E_{DNN}$ are the values of $\Delta\rho$ and $\Delta E$ predicted by the DNN for the $i$-th molecule, and $w_E$ is a coefficient defining a relative weight of the electron density and energy discrepancies in the overall estimate of the performance of the DNN. The matrix L1 norm in the loss function in eq. (7) is interpreted as follows:

$$\left\| a(i) \right\|_1 = \sum_k \left| a_k(i) \right|, \quad (8)$$

where $k$ runs over all points on a $64 \times 64 \times 64$ grid.

We split the QM9 database into training, validation and testing subsets based on the indices of the molecules in the database. The testing set is formed by molecules with indices matching the mask "???0??", where "?" stands for any digit (0 to 9), the validation set has the mask "???1??", and the testing set has the mask "???[2-9]??", where "[2-9]" stands for any digit from 2 to 9. In this way, ~10% of the database (~13K molecules) are reserved for validation, and the same number of molecules for testing. The masks are chosen in this way to ensure that all three subsets evenly cover the whole dataset, and that the first 99 molecules from the database, for some of which high-level quantum chemical computations are available, are not included into the training or even validation subsets, so that we can get an unbiased estimate of the performance of the DNN relative to quantum chemical methods more precise than DFT. This preprint reports the results of training a DNN on a subset of QM9. In a forthcoming publication, we are planning to present the results of training, validation and testing on the whole QM9 database, or at least its larger part. Learning curves for other models on the QM9 dataset typically demonstrate a significant drop in the error by ~35-50 thousand molecules in training sets, though further improvement with larger training sets still takes place.[26,32,33] The DNN reported in this preprint was trained on 38 268 molecules (QM9 indices matching the mask "???[2-9]??", in the range from 200 to 95 918, with omissions), and validated on 9 537 molecules (QM9 indices matching the mask "???1??", in the range from 100 to 95 198, with omissions). We have not analyzed the performance of the DNN on molecules from the testing set; this will be done in a final publication with a final version of a DNN trained on the whole training set of the QM9 database.

The DNN reported here was trained in two stages with different values of the learning rate. First, is was set to $2 \cdot 10^{-4}$, and eight epochs of training were performed. During each epoch, all molecules in the training set, randomly sorted, were processed in minibatches of 16 molecules (such that each molecule was used once and only once during each epoch). After that, the learning rate was decreased to $2 \cdot 10^{-5}$, and ten more epochs of training were carried out. The value of $w_E$ was set to a value with which the contribution of the energy term to the total loss function $L$ was ~10% by the end of the first epoch (as measured on the validation set), and stayed at this level during the whole training process. An attempt to increase $w_E$ by a factor of 10 did not lead to significant changes in the performance of the trained model in terms of either $\Delta\rho$ or $\Delta E$ prediction.

During training, the loss function on the training set, as expected, gradually decreased. The loss function computed on the validation set also tended to decrease, and stabilized by the end of training at a slightly higher level that the loss function for the training set (Fig. 7a). The plot reports relative values $Lrel$ of the loss function, computed as $L$ defined above divided by the value of this function for a DNN that would predict all values of $\Delta\rho$ and $\Delta E$ to be zeros:

$$Lrel = \frac{\sum_{i \in \text{training set}} \|\Delta\rho_{DNN}(\rho_{HF}(i)) - \Delta\rho(i)\|_1 + w_E \sum_{i \in \text{training set}} |\Delta E_{DNN}(\rho_{HF}(i)) - \Delta E(i)|}{\sum_{i \in \text{training set}} \|\Delta\rho(i)\|_1 + w_E \sum_{i \in \text{training set}} |\Delta E(i)|}, \quad (9)$$

which physically corresponds to the statement that the 'exact' (computed by DFT, in this case) electron density and energy equal those computed with HF/cc-VDZ. Hence, the value of $Lrel = 0$ corresponds to perfect prediction of the ground truth densities and energies by the DNN, while $Lrel = 1$ corresponds to the model being as bad as HF/cc-VDZ. The values of $Lrel$ for the validation set were computed at the end of each epoch (with the enumeration of epochs starting from 1) using the same state of the DNN (after training on all minibatches in the corresponding epoch) and all molecules from the validation set. The values of $Lrel$ for the training set were computed differently: while minibatches were consecutively treated during the corresponding epoch, the DNN was updated after each minibatch, and the contributions to $L$ for the molecules in the current minibatch were computed with the current DNN. Therefore, the overall values of $L$ and $Lrel$ reported for the training set represent different states of the DNN being trained during a certain epoch. Respectively, the values of $Lrel$ are shown in Fig. 7a in the middle between the indices of the previous and next epochs.

The performance of the DNN separately for $\Delta\rho$ and $\Delta E$ prediction shows different dynamics during training. The plot for the relative performance $L_{\Delta\rho}rel$ of the DNN in predicting $\Delta\rho$, defined as follows:

$$L_{\Delta\rho}rel = \frac{\sum_{i \in \text{training set}} \|\Delta\rho_{DNN}(\rho_{HF}(i)) - \Delta\rho(i)\|_1}{\sum_{i \in \text{training set}} \|\Delta\rho(i)\|_1}, \quad (10)$$

demonstrates the dynamics similar to that of $Lrel$ on the training and validation sets (Fig. 7b). By the end of training, $L_{\Delta\rho}rel$ on the validation set decreases to 0.128 and stabilizes at this level for 2-4 last training epochs. Therefore, predictions of the DNN for the electron density are much closer to the ground truth values

**Fig. 6**. Architecture of the reported DNN includes a U-Net part and a fork into density and energy prediction paths, with multiple three-dimensional (3D) convolution and 3D deconvolution operations. First, the input approximate electron density $\rho_{HF}$ undergoes a tanh transformation to ensure that information on chemical bonding is not dwarfed by atomic core densities, and artifacts of a discrete representation of the density near atomic cores are removed. Next, in the U-Net part (*light blue background*), 10 hidden layers deeply process the information by five 3D convolution and five 3D transposed convolution ('deconvolution') operations, with a horizontal flow of information between the layers of the same grid sizes to ensure high spatial resolution of the output $\Delta\rho$. The same set of highly processed characteristics of the system is used to compute, after two 3D convolution layers, the correction to the energy $\Delta E$ (*light yellow background*). This design of the DNN should push the energy predictions to be more based on high-level quantum mechanical properties of the modeled molecular system, as opposed to curve-fitting. The information processing operations are shown with different arrows labeled in the inset (b-e): (b) 3d convolution with a subsequent rectified linear unit (ReLu) activation; (c) transposed 3d convolution ('deconvolution') of a lower resolution cube, concatenation with a higher resolution cube, 3d convolution of the result, nonlinear ReLu activation; (d) 3d convolution with a subsequent ReLu activation; (e) 3d convolution without activation. ▶

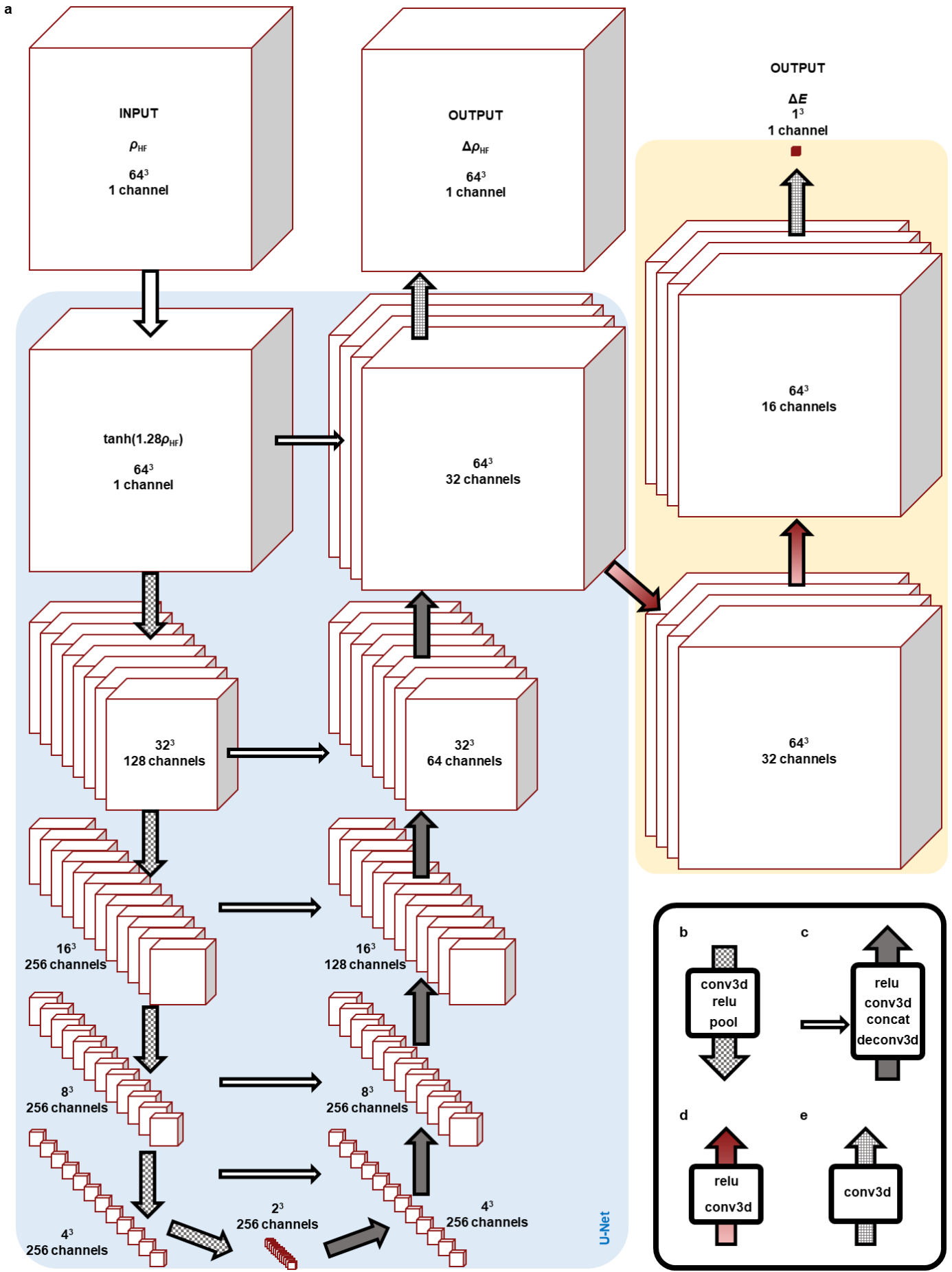
22

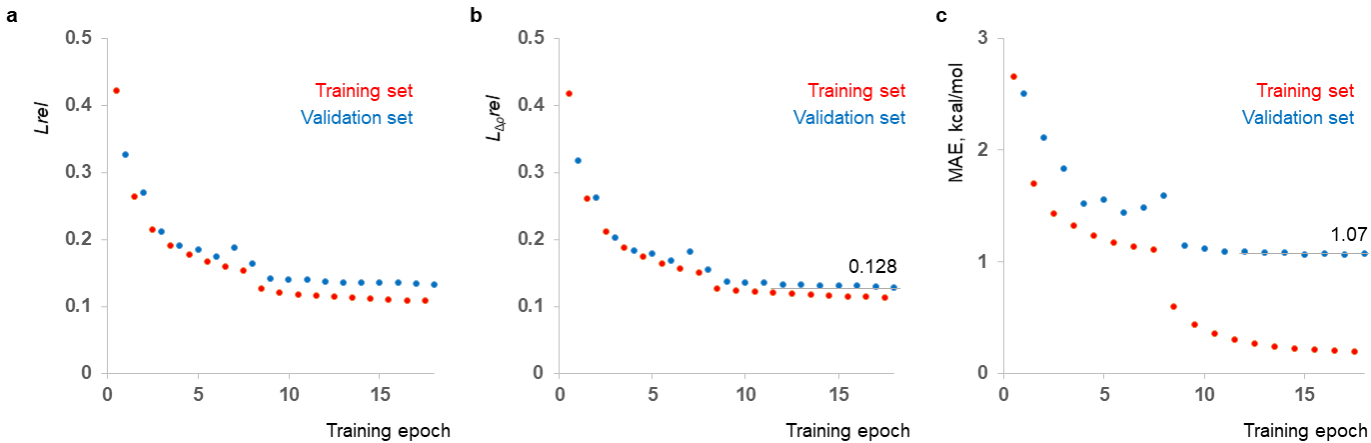

**Fig. 7**. Learning curves for the total loss function of the DNN (a) and its performance in predicting electron densities (b) and energies (c) demonstrate that after training the DNN converges to a state capable of predicting both electron densities and energies much closer to the DFT values than the HF method with cc-VDZ basis set. Panel (a) reports the values of the loss function $L$ relative to its value for zero output of the DNN $L_{rel}$, eq. (9). $L_{rel} = 0$ corresponds to perfect prediction of the DFT densities and energies by the DNN, while $L_{rel} = 1$ corresponds to the model being as bad as HF/cc-VDZ. Panel (b) reports a similar relative value for $\Delta\rho$ predictions only, $L_{\Delta\rho}rel$ [0 for perfect predictions of densities by the DNN, 1 for predictions as bad as HF/cc-VDZ, eq. (10)]. Panel (c) reports mean average error (MAE) in predicted energies, eq. (11), in kcal/mol. With zero output of the DNN, MAE would be 5.42 kcal/mol. ML models for energies for the QM9 dataset reported in the literature have MAE on the order of ~1 kcal/mol (Table 4).

computed from DFT ($L_{\Delta\rho}rel = 0$) than to the input density computed from HF/cc-VDZ ($L_{\Delta\rho}rel = 1$). The difference of $L_{\Delta\rho}rel$ over the training and validation sets is minor, implying that there is no significant overfitting of the DNN on the electron density data, and allowing us to expect that it may be transferable to other molecules similar to those in the QM9 datasets.

As for the DNN performance in predicting energies, we report the values of mean absolute error (MAE) in energies widely used in the literature on quantum chemical benchmarking:

$$MAE = \frac{1}{\|S\|} \sum_{i \in S} |\Delta E_{DNN}(\rho_{HF}(i)) - \Delta E(i)|, \quad (11)$$

where $S$ is a set of molecules (training or validation), and $\|S\|$ is the number of molecules in set $S$. Evidently, the energy term in the total loss function $L$, eq. (7), is proportional to MAE. The mean average value of the ground truth values $\Delta E(i)$ equals 5.42 kcal/mol. During training, MAE computed over the validation set decreases to 1.07 kcal/mol, staying stable at this level over the last 4-6 training epochs (Fig. 7c). Unlike the case of the electron densities, the performance of the DNN on the training data is much better, reaching MAE of 0.19 kcal/mol by the end of training. Therefore, the DNN overfits on the energy data, but not on the electron density data, presumably due to a much smaller set of datapoints for training (roughly speaking, 1 number for the energy vs. 64 × 64 × 64 = 262K numbers for the electron density per molecule). We attempted to resolve the problem of overfitting on the energy data by changing the architecture of the network (reducing the number of layers in the path leading to the energy values), but this deteriorated the performance of the model on the validation test set. The value of MAE demonstrated by the DNN on the validation set (1.07 kcal/mol) is on par with MAE values of other ML models over the QM9 dataset reported in the literature (Table 4).

The results on energy prediction reported above refer to the energy computed with PBE0/pcS-3. We have also tried to use the energies of molecules provided in the QM9 dataset, which were computed with other DFT functional and basis set, namely B3LYP/6-31G(2df,p).[43] With the same DNN architecture and training schedule, the MAE on validation set decreased only to 1.49 kcal/mol. A worse performance of the DNN on the combination of electron densities and energies from different sources [from PBE0/pcS-3 and B3LYP/6-31G(2df,p), respectively] might be caused by the architecture of the DNN, where calculations of the density and energy fork late in the network, and are based on a shared set of high-level features of the molecule. We speculate that the DNN might have learned some features characteristic of the PBE0/pcS-3 solutions, as opposed to the exact solutions, and the use of energies from another approximate method might cause the loss of accuracy of the model.

**Data availability**

The fchk files for all QM9 molecules computed with PBE0/pcS-3 and HF/pcS-3, as well as a file with the corresponding energies, are being uploaded to Stanford Digital Repository and will be available shortly by the following link: https://purl.stanford.edu/kf921gd3855 The code for the DNN and the cube files used for training will be made publically available after the paper is accepted.

## Conflict of Interests

V.S.P. is a consultant & SAB member of Schrodinger, LLC and Globavir, sits on the Board of Directors of Apeel Inc, Asimov Inc, BioAge Labs, Freenome Inc, Omada Health, Patient Ping, Rigetti Computing, and is a General Partner at Andreessen Horowitz.

## Acknowledgements

The authors thank other members of the Pande group, in particular, Dr. Joseph Gomes, for discussions of our results (which does not make them responsible in any possible way for the limitations and deficiencies of this work). We thank Stanford University and the Stanford Research Computing Center for providing computational resources (Sherlock and XStream clusters) and support that have contributed to these research results. XStream computational resource is supported by the National Science Foundation Major Research Instrumentation program (ACI-1429830). The Pande group is broadly supported by grants from the NIH (R01 GM062868 and U19 AI109662) as well as gift funds and contributions from Folding@home donors. We acknowledge the generous support of Dr. Anders G. Frøseth and Mr. Christian Sundt for the work on machine learning in the Pande group.